# Security and Privacy Management of IoT Using Quantum Computing

Jaydip Sen

Department of Data Science and Artificial Intelligence
Praxis Business School, Kolkata, INDIA
jaydip.sen@acm.org

**Abstract:** Quantum computing is increasingly recognized as a disruptive force in cybersecurity with significant implications for the security and privacy of the Internet of Things (IoT). Billions of IoT devices currently rely on traditional cryptographic methods, such as RSA and Elliptic Curve Cryptography (ECC), to protect sensitive data and enable secure communications. However, advances in quantum algorithms like Shor's and Grover's demonstrate the potential to break or weaken these widely deployed systems, thereby exposing IoT infrastructures to unprecedented risks. This chapter explores the challenges and opportunities that quantum computing introduces into the IoT ecosystem, where devices are often resource-constrained and operate under strict power and performance limitations. The discussion begins with an overview of classical cryptographic approaches and highlights their vulnerabilities to quantum attacks. It then examines the emerging field of *post-quantum cryptography* (PQC), focusing on algorithmic families such as lattice-based, code-based, and hash-based schemes, and analyzes their suitability for IoT deployments. The role of quantum-based techniques, particularly *Quantum Key Distribution* (QKD) and *Quantum Random Number Generators* (QRNGs), is also considered as complementary mechanisms for ensuring provable security and enhancing privacy in critical environments like smart cities and healthcare systems. Furthermore, the chapter addresses implementation challenges, including efficiency, salability, and interoperability, while emphasizing the importance of hybrid quantum-classical security architectures. By integrating current research insights with future directions, the chapter provides a comprehensive perspective on how quantum technologies will reshape IoT security frameworks and highlights strategies necessary for building a resilient, privacy-preserving, and quantum-safe IoT ecosystem.

## 4.1 Introduction

The Internet of Things (IoT) refers to a distributed system of interconnected physical devices embedded with sensors, software, and network connectivity that enables data collection and autonomous actuation [1-2]. The IoT has evolved into a foundational component of modern infrastructure [1-3]. Devices such as wearable health monitors, intelligent home appliances, and industrial automation systems exemplify the pervasive integration of IoT across personal, commercial, and civic domains [1, 4]. They are embedded in homes, workplaces, cities, and industries, creating networks of connected devices that collect, share, and act on data. This connectivity brings incredible convenience and new possibilities, but it also opens a plethora of security and privacy challenges [5].

One of the main challenges of IoT is its diversity and size. IoT devices come in many forms: tiny soil sensors, fitness trackers, traffic cameras, or factory equipment [1,6]. They use different communication methods and usually have limited power and computing abilities. Securing such a large and varied network is complex. Adding to the difficulty is the huge amount of data these devices collect. Much of it is personal or sensitive [7]. For example, fitness trackers collect health data, and smart meters track energy use in detail. If this data is stolen, it could lead to privacy violations, identity theft, or physical risks. Sadly, many IoT devices lack strong built-in security because manufacturers often focus more on cost and functionality than on protecting data [8].

Cryptography is the key to securing IoT communication. Traditional cryptographic methods, such as RSA and Elliptic Curve Cryptography (ECC), are based on computationally hard mathematical problems [9]. These methods have kept online banking and other systems safe for decades. However, quantum computing could change everything.

Quantum computers are very different from classical ones. They use quantum bits, or qubits, which can be in many states at once [10]. This lets quantum computers solve certain problems much faster. For example, Shor's algorithm can quickly factor large numbers, breaking RSA and ECC encryption regardless of the key size [11]. Grover's algorithm, meanwhile, weakens symmetric encryption like Advanced Encryption Standard (AES) by halving its effective strength [12]. So, AES-256 would offer only about 128-bit security against a quantum attack.

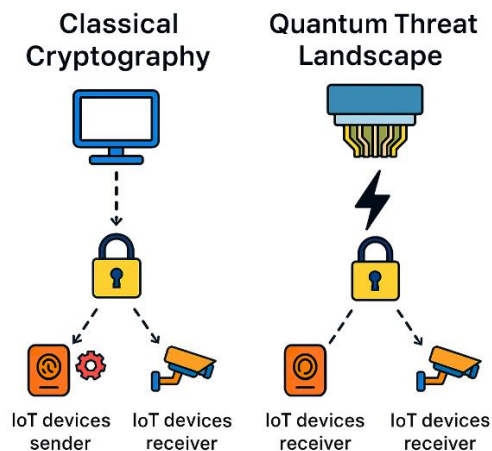

**Fig 4.1**: From classical ciphers to quantum storms: Visualizing the growing security divide in the IoT era

This means that once quantum computers become powerful enough, current cryptographic systems could fail. IoT devices being deployed today may still be in use when quantum computers arrive. If those devices are not quantum-safe, their data could be exposed. Worse, attackers could record encrypted data now and wait until they can decrypt it using quantum machines later, a method known as *store now, decrypt later* [13]. Fig. 4.1 depicts the security threats posed by the quantum computing paradigm.

To address this, a new field called *post-quantum cryptography* (PQC) is developing cryptographic systems that can resist quantum attacks but still work on classical devices [14]. These algorithms look promising but often demand more memory and processing power than many IoT devices can handle [15-16]. Making PQC efficient and lightweight enough for IoT is a major area of research.

Alongside PQC, quantum cryptography offers an exciting alternative approach. Using the fundamental laws of physics, techniques like Quantum Key Distribution (QKD) enable two parties to create secure encryption keys that cannot be intercepted without detection [17]. However, the specialized hardware and infrastructure needed for QKD currently limits its deployment, especially in the low-power, diverse world of the IoT [18].

The risks are even greater when we look at IoT's role in smart cities, healthcare, and national infrastructure [19]. A security breach in these systems could cause traffic chaos, harm patients, or disrupt critical services. As quantum computing gets stronger, securing IoT is not just a technical goal; it is vital for public safety and trust [20].

This chapter adopts a tutorial approach to guide readers through the emerging landscape of IoT security in the quantum computing era. Rather than presenting a novel algorithm or scheme, it aims to provide a structured and accessible explanation of the issues, solutions, and future directions. The discussion begins with a review of classical cryptographic methods currently used in IoT systems and their mathematical foundations, followed by an analysis of how quantum algorithms compromise their security guarantees. The chapter then introduces post-quantum cryptographic schemes, emphasizing their potential and limitations in IoT contexts. Quantum-based techniques such as QKD and Quantum Random Number Generators (QRNGs) are also covered, particularly in relation to complex environments like smart cities, where privacy and resilience are paramount.

The key contributions of this chapter are as follows. First, it provides a comprehensive classical cryptographic techniques relevant to IoT, including symmetric encryption, asymmetric encryption, and hash functions, along with their strengths and limitations, Second, It explains, in a non-technical yet rigorous manner, the fundamental principles of quantum computing and the mechanisms of algorithms like Shor's and Grover's that directly threaten contemporary cryptographic systems. Third, it introduces readers to the major families of post-quantum cryptography, highlighting their security assumptions, practical challenges, and applicability to resource-constrained IoT environments. Fourth, it discusses quantum-based methods, particularly QKD and QRNGs, and evaluates their role in securing IoT communications in critical infrastructures such as healthcare and smart cities. Finally, It presents a synthesis of existing research, implementation challenges, and future opportunities, serving as an entry point for researchers, practitioners, and students seeking to understand this rapidly evolving field.

The rest of the chapter is organized as follows. Section 4.2 begins with a brief overview of classical cryptographic techniques: symmetric key systems, public key infrastructures, and hash functions, highlighting their role in securing modern IoT ecosystems. Section 4.3 provides a detailed assessment of quantum computing capabilities, with a focus on the resource requirements and implications of quantum algorithms such as Shor's and Gover's for breaking classical cryptosystems. Building on this, Section 4.4 introduces the core principles and major families of PQC algorithms, emphasizing their relevance and applicability to resource-constrained IoT devices. Section 4.5 takes the discussion further by focusing on urban-scale deployments of quantum cryptography and its integration into smart city infrastructure, including hybrid quantum-classical security architecture and practical implementation models. Section 4.6 explores privacy management strategies in a post-quantum world, incorporating regulatory compliance, federated learning, and quantum-resilient privacy-preserving mechanisms. Finally, Section 4.7 concludes the chapter highlighting some future research directions.

## 4.2 Cryptographic Algorithms

This section breaks down three key types of cryptographic algorithms: symmetric cryptography, asymmetric cryptography, and hash functions. Each plays a distinct role in protecting data, but as quantum computing evolves, their effectiveness is being called into question [4, 6]. Understanding how these algorithms work, where they are used, and why they may no longer be secure against quantum attacks is crucial for building the next generation of IoT security. Fig. 4.2 shows how symmetric and asymmetric cryptographic models differ in their key usage mechanism.

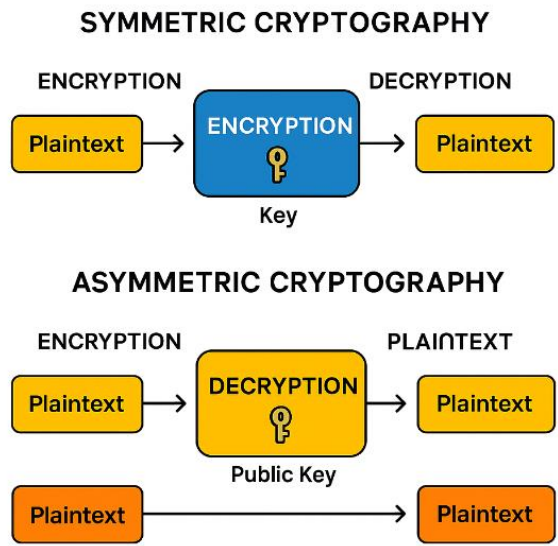

**Fig 4.2**: Symmetric vs. asymmetric cryptography showing shared-key and public-private key usage.

## 4.2.1 Symmetric Key Cryptography

Symmetric key cryptography remains the most fundamental techniques in securing digital communication. At the heart of symmetric key cryptography is the *single shared key* that the sender and the receiver use to encrypt and decrypt messages. This shared key must remain confidential [21]. If it is compromised, the security of all communication using the key is jeopardized.

*Core Principles and Workflow:* The basic principle of symmetric encryption is straightforward: the same key $K$ is used to convert readable data (plaintext $M$) into an unreadable form (ciphertext $C$) and vice versa. The encryption function $E$ and decryption function $D$ are mathematically related such that $C = E_K(M)$ and $M = D_K(C)$, where the encryption key ($E_K$) and the decryption key ($D_K$) are identical. The correctness property demands that applying the decryption function to the ciphertext recovers the original message (i.e., plain text).

$$D_K\big(E_K(M)\big) = M \tag{1}$$

The relationship in (1) ensures the integrity of symmetric encryption and ensures the messages can be securely transmitted and reliably recovered.

### 4.2.1.1 Types of Symmetric Key Algorithms

Symmetric key cryptographic algorithms primarily fall into two categories: (i) Block Ciphers and (ii) Stream Ciphers.

*Block Ciphers* encrypt fixed-size blocks of data (commonly 64 or 128 bits). Each block is transformed independently or in combination with others through multiple rounds of substitutions and permutations to produce ciphertext blocks. A popular block cipher is the AES, standardized by National Institute of Standards and Technology (NIST) in 2001 [22]. AES has become the *de facto* symmetric encryption method due to its strong security and efficiency. Block ciphers such as AES operate on fixed-size blocks of plaintext (e.g., 128 bits). However, most real-world messages are longer and require mechanisms to process variable-length data. This is achieved using modes of operation, which define how blocks are linked and processed during encryption. Three common modes are: (i) *electronic codebook* (ECB), (ii) *cipher block chaining* (CBC), and (iii) *Galois/Counter mode* (GCM). ECB encrypts each block independently, which is fast but leaks data patterns and is generally not recommended for secure applications [23]. In CBC, each plaintext block is XORed with the previous ciphertext block before encryption, improving diffusion but requiring an *initialization vector* (IV) and being vulnerable to padding oracle attacks if not used carefully [24]. GCM combines counter mode encryption with a polynomial-based authentication tag, making it well-suited for high-performance and secure communication [25]. In IoT systems, GCM is preferred in protocols like TLS [26] due to its lightweight authentication and parallelizability, which are valuable for constrained devices.

*Stream Ciphers* generate a pseudorandom stream of bits (keystream) which is combined with plaintext bits, typically via the XOR operation [27]. This method encrypts data bit-by-bit or byte-by-byte, making it suitable for applications requiring real-time encryption, such as voice or video streaming. RC4 [28] is a classic example of stream ciphers, though newer, more secure stream ciphers have been developed.

### 4.2.1.2 Mathematical Foundations of Symmetric Key Algorithms

The security and operation of symmetric key algorithms rely heavily on discrete mathematics, including permutations, substitutions, finite fields, and modular arithmetic.

*Permutation and Substitution*: Block ciphers often use these two operations in a layered fashion. Permutations reorder bits or bytes within a block, while substitutions replace bits or bytes with others according to a predefined mapping (e.g., S-boxes in AES). These steps introduce *confusion* and *diffusion*, two essential properties to thwart cryptanalysis. While confusion obscures the relationship between the key and ciphertext, diffusion spreads the influence of a single plaintext bit over many ciphertext bits, making patterns less discernible.

*Finite Field Arithmetic*: AES, for instance, operates in the field $GF(2^8)$, where bytes are treated as elements of an algebraic structure with 256 elements. Operations like multiplication and addition in this field obey specific rules crucial for AES's substitution and mixing steps. For example, *MixColumns* step in AES can be expressed as a matrix multiplication over $GF(2^8)$:

$$MixColumns(S) = M \times S \qquad (2)$$

In (2), $S$ is the state matrix representing the block, and $M$ is a fixed invertible matrix in $GF(2^8)$.

*Key Size and Security*: The length of the secret key $n$ is directly tied to the security level. The key space consists of $2^n$ possible keys. A brute-force attack, which involves trying every possible key, has an expected complexity of $2^{n-1}$ attempts on average. AES supports keys of 128, 102, and 256 bits. AES-256, for example, offers $2^{256}$ possible keys, making brute-force attacks computationally infeasible with current technology.

Fig. 4.3 depicts the schematic of the AES cryptosystem operations.

### 4.2.1.3 Applications of Symmetric Key Algorithms in IoT

IoT devices operate under resource constraints such as limited computational power, memory, and battery life. Symmetric cryptography is well-suited for such environments because of its computational efficiency compared to asymmetric methods. Common IoT security frameworks use symmetric algorithms to encrypt sensor data, control messages, and firmware updates. Protocols like MQTT [29], CoAP [30], and DTLS [31] frequently employ symmetric encryption to protect data-in-transit. However, securely distributing and managing symmetric keys in large, dynamic IoT networks remains challenging. Pre-shared keys are vulnerable if devices are captured or tampered with, while key distribution protocols need to be lightweight yet secure.

Some emerging lightweight symmetric key cryptographic schemes suited for IoT environment are: PRESENT [32], SPECK [33], and SIMON [33].

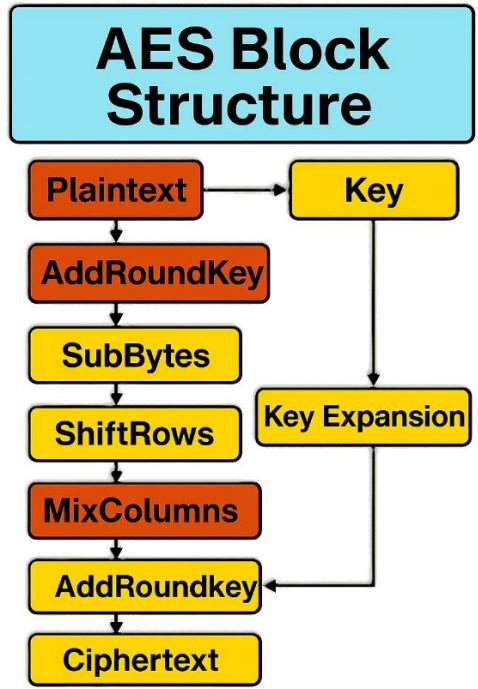

**Fig 4.3**: Structural flow of AES encryption showing substitution, permutation, and mixing steps.

### 4.2.2 Asymmetric (Public) Key Cryptography

In contrast to symmetric key cryptography, which relies on a single shared key, asymmetric key cryptography, also known as public key cryptography, employs a pair of mathematically related keys: a public key and a private key. This dual-key system addresses the critical issue of key distribution in large networks, enabling secure communication without requiring the exchange of secret keys. While the public key is openly distributed and used for encryption, the private key is kept confidential and used for decryption. Asymmetric cryptography includes many critical security protocols, including digital signatures, key exchange mechanisms, and secure communication channels.

*Fundamental Principles and Workflow*: Asymmetric encryption is based on mathematical problems that are computationally hard to reverse without specific knowledge of the private key [21]. The encryption and decryption functions are defined as follows.

*Encryption Function (E)*: Given a plaintext $M$ and a public key $P$, the ciphertext $C$ is computed as: $C = E_P(M)$.

*Decryption Function (D)*: Given the ciphertext $C$ and the private key $S$, the original plaintext $M$ is recovered as: $M = D_S(C)$.

The requirement for this encryption system is given by (3):

$$D_S(E_P(M)) = M \qquad (3)$$

Thus, even if the public key $P$ is known, without the private key $S$, it is computationally infeasible to derive the plaintext from the ciphertext.

*Mathematical Foundations and Algorithms*: Asymmetric encryption leverages complex mathematical structures, such as number theory, elliptic curves, and modular arithmetic, to establish secure key pairs. The most prominent algorithms include RSA, Diffie-Hellman, and ECC.

### 4.2.2.1 The RSA Algorithm

The RSA algorithm [34], named after Rivest, Shamir, and Addleman, is based on the mathematical hardness of the factorization problem for large composite numbers. The key steps of the RSA algorithm are illustrated in Fig 4.4. The steps are discussed in the following:

*Key Generation*: This step includes the following: (i) Select two large prime number $p$ and $q$. (ii) Compute their product $n = p * q$. (iii) Compute the Euler totient function $\Phi(n) = (p - 1) * (q - 1)$. (iv) Select an encryption exponent $e$, such that $1 < e < \Phi(n)$ and $GCD(e, \Phi(n)) = 1$. Note, *GCD* denotes the Greatest Common Divisor. (v) Calculate the decryption exponent $d$ as the multiplicative inverse of $e \bmod \Phi(n)$: $d * e \equiv 1 (mod\ \Phi(n))$. (vi) The public key is $(e, n)$, and the private key is $(d, n)$.

*Encryption and Decryption:* The encryption step computes the cyphertext $(C)$, given the plaintext $(M)$ using: $C = M^e\ mod\ n$. The decryption step retrieves the plain text using: $M = C^d\ mod\ n$. The security of RSA is based on the computational complexity of factoring the large composite number $n$ into its two prime factors $p$ and $q$. The strength of RSA increases exponentially with the key size. Common key sizes include 2048 and 4096 bits [34].

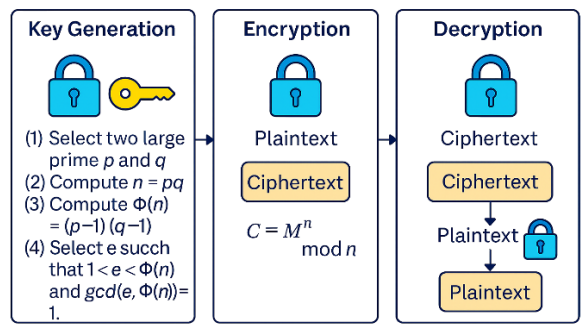

**Fig 4.4**: Workflow of RSA cryptosystem showing key generation, encryption, and decryption.

### 4.2.2.2 The Diffie-Hellman Key Exchange Protocol

The Diffie-Hellman Key Exchange (DHKE) Protocol is a cryptographic technique used to securely establish a shared secret key between two parties communicating over an unsecured channel [35]. The security of the protocol is based on the computational hardness of the *discrete logarithm problem*, which becomes intractable when large prime numbers are used.

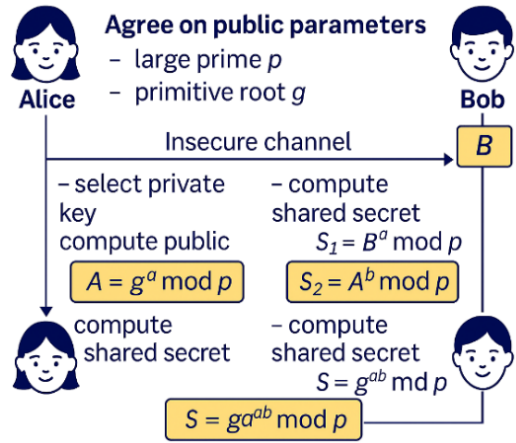

**Fig 4.5**: Diffie-Hellman key exchange showing parameter setup, key generation, and shared secret derivation.

Fig 4.5 illustrates the DKKE protocol. The steps involved in the DHKE protocols are as follows. First, two persons Alice and Bob agree on two public parameters: a large prime $p$ and a corresponding *primitive root* modulo $p$, denoted as $g$, such that $1 < g < p$. Following this, each party independently selects a secret *private key*. Alice chooses a random integer $a$, where $1 < a < p$, and Bob selects a random integer $b$, satisfying $1 < b < q$. Using their shared base $g$ and modulus $p$, each party computes their respective *public keys*. Alice computes her public key as $A = g^a \bmod p$, while Bob computes his public key as $B = g^b \bmod p$. These public keys, $A$ and $B$, are then exchanged over the unsecure channel.

Upon receiving the other party's public key, both Alice and Bob proceed to compute the *shared secret key*. Alice uses Bob's key $B$ along with her private key $a$ to compute $S_1 = B^a \bmod p$, while Bob uses Alice's public key $A$ and his private key $b$ to compute $S_2 = A^b \bmod p$. Due to the properties of modular exponentiation, both parties arrive at the same result: $S = S_1 = S_2 = g^{ab} \bmod p$. The shared secret $S$ is known only to Alice and Bob, and eavesdropper, despite knowing $p, g, A,$ and $B$, cannot feasibly compute $S$ without solving the discrete logarithm problem.

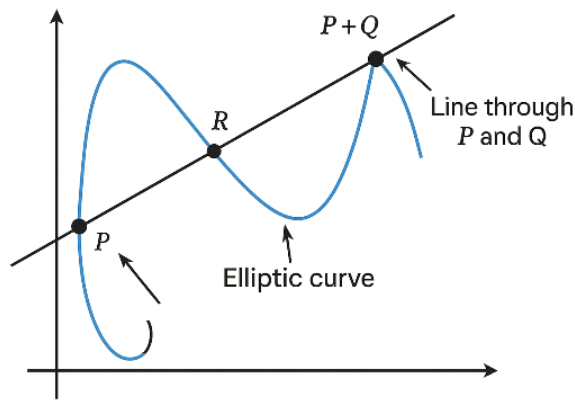

**Fig 4.6**: Point addition of two points P and Q in elliptic curve cryptography (ECC).

### 4.2.2.3 The Elliptic Curve Cryptography (ECC)

*Elliptic Curve Cryptography (ECC)*: ECC is a more modern asymmetric encryption method that offers equivalent security to RSA but with significantly smaller key sizes [36-37]. ECC is based on the mathematics of elliptic curves over finite fields [38]. A typical elliptic curve is defined in (4):

$$y^2 = x^3 + ax + b \tag{4}$$

In (4), $a$ and $b$ are constants and $x$, $y$ are points on the elliptic curve. The *key generation* and the encryption and decryption using of ECC are carried out as follows. The key generation step of ECC involves the following steps: (i) A base point $G$ on the curve is selected. (ii) A randomly chosen integer $d$ is selected as the private key. (iii) The public key $Q$ is derived using $Q = d * G$. In the encryption step, the plaintext is chosen as a point $P$ on the curve. The ciphertext is computed as a pair of points: $C = (kG, P + Q)$, where $k$ is a random integer. In the decryption step, the plaintext $P$ is computed using $P = (P + kQ) - d(kG)$. As depicted in Fig 4.6, the point addition operation in ECC forms the basis of ECC's group structure, where the sum of two points on the curve yields a third point through geometric reflection.

*Security Considerations of ECC*: ECC provides security even with smaller key sizes. A 256-bit ECC key is considered as secure as a 3072-bit RSA key [39]. This makes ECC particularly suitable for resource-constrained IoT devices.

### 4.2.2.4 Applications of Asymmetric Key Cryptography in IoT

In IoT systems, asymmetric cryptography is widely employed for secure key exchange, device authentication, and digital signatures. IoT devices use protocols like Elliptic Curve Diffie-Hellman (ECDH) for secure key exchange [40]. ECDH provides security with minimal key size, making it suitable for constrained devices. Elliptic Curve Digital Signature Algorithm (ECDSA) ensures data integrity and non-repudiation in IoT communication [41]. IoT systems often adopt hybrid encryption [42]. In these hybrid encryption models, asymmetric cryptography is used to securely exchange a *session-specific symmetric key*, which is then used to encrypt the actual data payload. This approach leverages the speed and efficiency of symmetric encryption for bulk data handling while maintaining secure key distribution through asymmetric means.

### 4.2.3 Hash Functions

Hash functions are a foundational component of modern cryptography and cybersecurity. Unlike encryption algorithms that transform plaintext into ciphertext and are reversible (given the correct key), cryptographic hash functions are one-way mathematical functions. They take an input (or message) of arbitrary length and return a fixed-length output, known as the *hash value* or *message digest*. Hash functions play a central role in ensuring data integrity, generating digital signatures, and verifying message authenticity, all critical requirements for secure communication in IoT environments.

*Fundamental Properties of Cryptographic Hash Functions*: A secure cryptographic hash function $H$ must satisfy the following properties [43]: (i) deterministic, (ii) pre-image resistant (one-way trapdoor), (iii) second pre-image resistant, (iv) collision resistant, (v) sensitive to avalanche effect. The deterministic property implies that the same input to a has function always produce the same output. In other words, $H(m_1) = H(m_2) \Leftrightarrow m_1 = m_2$. The one-way trapdoor property implies that given a hash value $h$, it should be computationally infeasible to find an input $m$ such that: $H(m) = h$. The second pre-image resistant property implies given an input $m_1$, it should be hard to find another input $m_2 \neq m_1$ such that $H(m_1) = H(m_2)$. The collision resistant property indicates that it should be infeasible to find two distinct inputs $m_1 \neq m_2$ such that: $H(m_1) = H(m_2)$. The avalanche effect implies that a small change in the input (even one bit) should significantly change the output hash value.

*Mathematical Formulation*: Let $m \in \{0,1\}^*$ be a message of arbitrary length and let $H: \{0,1\}^* \rightarrow \{0,1\}^n$ be the hash function that outputs an $n$-bit hash: $h = H(m)$. Here, $m$ is the input message (e.g., a data packet, file, or password), $h$ is the resulting message digest, and $n$ is the fixed output size. Unlike encryption, hash functions do not require any keys.

*Some Popular Cryptographic Hash Functions*: Over the years, several hash functions have been developed, standardized, and, in some cases, deprecated due to emerging cryptanalytic attacks. Some of the well-known hash functions are briefly discussed in the following.

### 4.2.3.1 The Message Digest 5 Hash Algorithm

One of the earliest hash functions that achieved widespread adoption is the Message Digest 5 (MD-5) [44]. This hash function was designed by Ronald Rivest in 1991. It outputs a 128-bits hash value and processes data in blocks of 512 bits using a *Merkel-Damgard* construction [45]. Despite its historical popularity, MD-5 has been rendered cryptographically insecure. As early as 2004, researchers demonstrated that it is vulnerable to collision attacks [46]. In a collision attack, two distinct inputs could produce the same hash output. Consequently, MD-5 is now considered obsolete for security-sensitive applications.

### 4.2.3.2 The Secure Hash Algorithm 1 (SHA-1)

The SHA-1 algorithm was developed by the U.S. National Security Agency (NSA) and published as a federal standard by NIST in 1995 [47]. It produces a 160-bit digest and follows a *Merkle-Damgard* construction. Although SHA-1 was considered secure in early years, theoretical attacks emerged in 2005, casting doubt on its collision resistance [48]. In 2017, Google, in collaboration with CWI Amsterdam, publicly demonstrated a practical collision using what they termed the *SHAttered* attack [49]. This attack exploited structural weaknesses in the compression function of SHA-1 to produce two distinct PDF files with identical hash values. This effectively deprecated SHA-1 for all cryptographic purposes, including its use in Transport Layer Security (TLS) [50] and digital certificates [51].

### 4.2.3.3 The Secure Hash Algorithm 2 (SHA-2)

In response to the vulnerabilities of SHA-1 protocol, NIST published the SHA-2 family in 2001 [52]. This family includes several variants differentiated by their output lengths: SHA-224, SHA-256, SHA-384, and SHA-512, among others. The SHA-2 algorithms maintain the Merkle-Damgard framework but introduce stronger logical functions and larger word sizes to improve resistance to differential and collision-based attacks. Mathematically, SHA-256 takes an input message $m$ and produces a 256-bit output. SHA-2 is widely used today in critical infrastructures, including blockchain, digital signatures, secure boot processes, and TLS.

### 4.2.3.4 The Secure Hash Algorithm 3 (SHA-3)

Recognizing the structural similarities and potential limitations of the Merkle-Damgard construction employed by both SHA-1 and SHA-2, NIST initiated a public competition to develop a new hash standard. This led to the selection of the Keccak algorithm [53] as the basis for SHA-3, which was finalized in 2015 [54]. Unlike its predecessors, SHA-3 utilizes a fundamentally different design known as the *sponge* construction. This approach separates the hash function into two phases: (i) *absorption* and (ii) *squeezing*. During absorption, the input message is XORed into a subset of the internal state, which is 1600 bits in total, and then transformed using a permutation function. During the squeezing phase, the output bits are extracted from the same subset of state. The strength of the function depends on the division of the internal state into a "rate" (denoted as $r$) and "capacity" ($c$), satisfying the relation: $r + c = 1600$. Larger capacity provides stronger security against collision and preimage attacks, while a higher rate allows for faster processing. SHA-3 variants include SHA3-224, SHA3-256, SHA3-384, and SHA3-512. SHA-3 also supports extendable-output functions (XOFs) [54], namely SHAKE128 [55] and SHAKE256 [55], which allow users to generate digests of arbitrary length. These XOFs are increasingly valuable in key derivation functions and PQC protocols.

A comparative overview of widely known hash functions and their cryptographic properties in presented in Table 4.1.

**Table 4.1:** A comparative overview of widely known hash functions and their cryptographic properties

| Feature | Protocol | | | |
|---|---|---|---|---|
| | MD 5 | SHA-1 | SHA-2 (256) | SHA-3 (256) |
| Output Size (bits) | 128 | 160 | 256 | 256 |
| Structure | Merkle-Damgard | Merkle-Damgard | Merkle-Damgard | Sponge |
| Collision Resistance | Broken | Broken | Secure | Secure |
| Preimage Resistance | Weak | Weak | Strong | Strong |
| Length Extension Attack | Yes | Yes | Yes (HMAC-safe) | No |
| Performance | Fast | Moderate | Good | Moderate |
| Usage | None | Deprecated | Widely Used | Emerging |

### 4.2.3.5 Security of Cryptographic Hash Functions

The security of cryptographic hash functions is generally assessed based on its preimage resistance, (ii) second preimage resistance, and collision resistance. While MD-5 and SHA-1 have failed to uphold collision resistance under contemporary attack models, SHA-2 and SHA-3 remain secure against all known practical attacks and are recommended for use in secure systems.

**Table 4.2:** Comparison of symmetric key cryptography, asymmetric key cryptography , and cryptographic hash functions in the context of IoT security.

| Feature/Property | Symmetric Key Crypto | Asymmetric Key Crypto | Crypto Hash Functions |
|---|---|---|---|
| Typical Algorithms | AES, ChaCha20, PRESENT | RSA, ECC, Diffie-Hellman | SHA-2, SHA-3, BLAKE3 |
| Key Size (Typical) | 128-256 bits | 2048-4096 bits (RSA), 256 bits (ECC) | Not Applicable (No Key Required) |
| Speed / Efficiency | High (fast, low computational cost) | Low (slow on constrained devices) | Very High (one-way and efficient) |
| Use Cases in the IoT | Bulk encryption, firmware protection | Secure key exchange, device authentication | Data integrity, password verification |
| Reversibility | Reversible (with key) | Reversible (with private key) | Irreversible |
| Quantum Threat | Quadratic threat | Exponential threat (Shor's algorithm) | Minor to moderate (depends on construction) |
| Security Basis | Key secrecy | Hard mathematical problems (e.g., factoring ECDLP) | Collision and preimage resistance |
| Implementation Complexity | Low | High | Low |
| Resource Requirements | Low | High | Low |
| Standardization | FIPS-197 (AES), ISO/IEC 18033 | FIPS-186-4 (RSA, DSA, ECDSA) | FIPS-180-4 (SHA-2), FIPS-202 (SHA-3) |

In many IoT applications, integrity and authentication are enforced using *Hash-based Message Authentication Code* (HMAC) [56-57]. HMAC combines a cryptographic hash function with a secret key to produce a *message authentication code* (MAC), which ensures both data integrity and authenticity. Unlike raw hash function, HMAC is designed to be secure even when the underlying hash function is vulnerable to length-extension attacks. HMAC-SHA 256, for example, is commonly used in MQTT [58] and CoAP [59] protocols to authenticate messages and present tampering in low-power or latency-sensitive IoT environments.

Table 4.2 summarizes key differences between symmetric, asymmetric, and hash-based cryptographic methods in the context of IoT security and post-quantum considerations.

Section 4.4 provides insights into hash-based digital signatures schemes like SPHINCS+ and their role in building quantum-resilient authentication mechanisms. Further, privacy preserving architecture leveraging quantum techniques are elaborated in Section 4.6.

## 4.3 Size Estimation of Quantum Computers

Estimating the size, error tolerance, and execution complexity of quantum computers necessary to break existing cryptographic algorithms is central to understanding the urgency of adopting PQC. Classical encryption methods such as RSA and ECC are secure under current computational capabilities because they rely on the intractability of integer factorization and discrete logarithms. However, quantum algorithms like Shor's and Grover's present a paradigm shift. Shor's algorithm offers exponential speedup for problems like integer factorization and discrete logarithms, reducing their complexity from sub-exponential (in classical schemes) to polynomial time $O((logN)^3)$[60]. Grover's algorithm, on the other hand, reduces brute-force search from $O(2^n)$ to $O(2^{\frac{n}{2}})$[12]. However, it provides no exponential speedup for structural cryptanalytic attacks (e.g., differential, or algebraic attacks), which remain largely unaffected. Shor's and Grover's algorithms redefine the boundaries of what is computationally feasible.

To assess the real-world threat posed by quantum computers, it is crucial to quantify the quantum resources required to execute such algorithms at scale. This includes estimating the number of logical qubits, the required physical qubits for error correction, circuit depth, gate fidelity, and the coherence time over which quantum states must remain stable. Such estimates provide a benchmark for when current cryptographic systems may become vulnerable, forming the technical basis for the shift of PQC explored in Section 4.4.

### 4.3.1 Logical vs Physical Qubits

A *logical qubit* is a fully error-corrected qubit used to perform reliable quantum computations. A *physical qubit*, by contrast, is the actual physical system (e.g., a trapped ion or superconducting loop) that suffers from noise and decoherence. Since quantum information is highly susceptible to errors from environmental noise and imperfect gate operations, logical qubits must be constructed using multiple physical qubits through *quantum error correction* (QEC).

One of the most practical QEC techniques is the *surface code*, which encodes one logical qubit to a two-dimensional lattice of physical qubits [61]. The number of physical qubits $N_p$ required construct one logical qubit using the surface code grows approximately based on: $N_p = d^2$, where, $d$ is the code distance, determined by the desired fault-tolerance threshold. The code distance $d$ reflects the minimum number of errors needed to corrupt a logical qubit, with lager $d$ yielding greater fault-tolerance. In practical implementations $N_p$ ranges between 1000 to 5000 physical qubits per logical qubit depending on fault-tolerance targets.

### 4.3.2 Quantum Resource Estimation for RSA and ECC

Shor's algorithm poses a critical threat to widely used asymmetric cryptographic schemes, particularly RSA and ECC. The security of RSS relies of the computational difficulty of factoring large semiprime prime integers, while ECC is based on the hardness of solving the Elliptic Curve Discrete Logarithm Problem (ECDLP). Both problems are considered intractable for present-day computers at cryptographic key sizes. However, Shor's algorithm reduces their complexity to polynomial time, specifically $O((logN)^3)$, thereby rendering both RSA and ECC insecure.

Although ECC achieves comparable classical security with significantly smaller key sizes compared to RSA, the underlying mathematical hardness assumptions are equally vulnerable to quantum attacks. In quantum terms, the computational effort required to solve ECDLP is comparable to that needed for factoring RSA moduli or equivalent classical strength. Thus, quantum resource estimates for breaking RSA-2048 and ECC-256 both fall within the same general range of qubit requirements [61].

Practical execution of Shor's algorithm requires more than algorithmic formulation. It depends on robust fault-tolerant architecture. For both RSA and ECC, resource estimation involves compiling the quantum circuit into a series of fault-tolerant gates (e.g., Clifford + T gates), managing quantum Fourier transforms, and implementing modular arithmetic routines with high fidelity. The total number of logical qubits required scales linearly with the key size, while the number of physical qubits grows super-linearly due to quantum correction overhead. Moreover, modular exponentiation and inverse operations for large primes (used in RSA) are highly gate intensive. These operations increase the total circuit depth and pose a significant challenge.

These architectural requirements make resource estimation a critical step in post-quantum risk assessment. Quantum algorithms like Shor's are not practical today due to their enormous physical qubit demands and sensitivity to error rates. However, once fault-tolerant quantum hardware becomes scalable, asymmetric cryptographic schemes could be broken with high probability. This real threat necessitates urgent transition to quantum-safe alternatives.

### 4.3.3 Error Correction Overhead

Quantum error correction (QEC) is essential to overcome the fragility due to the decoherence and noise that affect quantum systems. One of the most practical QEC methods is the *surface code*. The surface code can operate reliably below a threshold gate error rate of approximately 1%, with some proposals pushing it down to 0.1%. The number of physical qubits per logical qubit increases quadratically with the code distance required to suppress logical errors [62]. A typical target logical error suitable for cryptographic application is $10^{-15}$. This demands physical qubits approximately estimated based on: $N_{physical} \approx 1000 * N_{logical}$. Here, $N_{physical}$ and $N_{logical}$ represent the number of physical and logical qubits, respectively. This scaling represents the most significant bottleneck for practical quantum cryptanalysis. The error rate per gate must be $10^{-3}$ or lower to be viable for large-scale cryptographic quantum computations.

### 4.3.4 Gate Depth and Circuit Complexity

The effectiveness of quantum algorithms also depends on the *circuit depth*. The circuit depth is defined as the longest chain of dependent quantum gate operations that directly affects the coherence time required [63]. Shor's algorithm involves intensive use of Toffoli gates (controlled-controlled-NOT) [64] and modular exponentiation [11]. For RSA-2048, the required gate depth is estimated to exceed $10^9$ operations [65]. *T*-gates, particularly Toffoli decompositions, dominate quantum runtime due to their cost in fault-tolerant implementations. A practical quantum computer must maintain coherence long enough to execute millions or even billions of gate operations.

To consolidate the resource demands across RSA and ECC cryptanalysis using Shor's algorithm, Table 3 summarizes estimated quantum requirements. These figures assume error-corrected quantum circuits using surface codes, low error rates, and gate fidelities suitable for scalable implementations. The column gate depth reflects the algorithmic complexity and coherence constraints. The figures suggest that RSA and ECC cryptographic key sizes would require several million high-fidelity qubits and hours of uninterrupted, error-corrected computation.

**Table 4.3:** Estimated quantum resource requirement for breaking RSA-2048 and ECC-256 using Shor's algorithm with quantum error correction (QEC)

| Algorithm | Logical Qubits | Physical Qubits | Gate Depth | Approx. Run Time |
|---|---|---|---|---|
| RSA-2048 | 4000 - 6000 | ~20 million | >$10^9$ gates | ~8 hours (with QEC) |
| ECC-256 (ECDLP) | 1000 - 2500 | ~10 million | ~$10^8$ - $10^9$ gates | ~4-5 hours |

As exhibited in Table 4.3, a recent study by Gidney and Ekerå (2023) estimated that factoring RSA-2048 using Shor's algorithm would require approximately 20 million physical qubits, assuming the presence of surface code with error correction and T-gate decomposition, and it would take about 8 hours of quantum computation assuming $10^6$ T-gate layers and high-fidelity two-qubit operations. This runtime assumes gate clock speeds in the order of tens of nanoseconds and massive concurrency in error-corrected quantum modules. These estimates represent optimistic lower bounds. Real-world implementations may require additional resources for quantum memory, synchronization, and control overhead.

### 4.3.5 Implications for PQC Timeline

Despite the immense quantum resources required, estimates suggest that cryptographically relevant quantum computers could become feasible within 10 to 30 years. This will depend heavily on advances in materials science, control engineering, and evolution of sophisticated error correction algorithms. This timeline may appear distant, but several factors demand immediate attention: First, the threat of *harvest now, decrypt later* should be handled. Encrypted data harvested today could be decrypted decades later, violating long-term confidentiality. Second, cryptographic systems in use today (e.g., smart grid, medical records) are expected to remain operational well into the quantum era. Hence, proactive security mechanisms must be in place before the quantum threats become a reality. Third, there is an urgent need for standardization. Organizations like NIST and ETSI are finalizing PQC standards, and transitioning early ensures smoother integration. In 2022, NIST announced the selection of CRYSTALS-Kyber and Dilithium for standardization, with final Federal Information Processing Standards (FIPS) publication expected by 2025 [66].

This technical landscape forms the basis for transitioning toward PQC, which is explored in detail in Section 4.4. Estimating the quantum resources required to break RSA and ECC highlights the long but narrowing window for PQC adoption [67]. Though millions of physical qubits and hours of coherent computation are currently required, the progress in quantum hardware makes the threat realistic in the medium term. The costs of inaction: compromised privacy, critical infrastructure exposure, and national security risks, are too high to delay proactive migration.

## 4.4 Transitioning to Quantum-Resilient Cryptography

As mentioned in Section 4.3, the advent of quantum computing presents a formidable challenge to the cryptographic foundations of modern digital communication. Algorithms such as Shor's and Grover's compromise the security of widely deployed schemes such as RSA, ECC, DSA, and symmetric-key systems. To address these threats, PQC seeks to design schemes based on problems believed to be hard even for quantum computers [68]. Unlike QKD, PQC can be deployed using classical networks and hardware, making it practical for today's digital systems. The following subsections examine four major families of PQC schemes: lattice-based, hash-based, code-based, and multivariate, each with distinct mathematical foundations, performance characteristics, and security guarantees.

### 4.4.1 Lattice-Based Cryptography

*Concepts and Motivation*: Lattice-based cryptography constructs cryptographic schemes using geometric structures in multi-dimensional spaces called lattices [69]. These lattices consist of regular arrangements of points in space. Solving certain problems on them, such as finding the shortest or closest point, is known to be computationally hard, even for quantum computers. Fig 4.7 illustrates how lattice-based cryptography relies on structured matrix operations and trapdoor functions to enable secure encryption and decryption processes.

*Mathematical Foundation*: A lattice $\mathcal{L}$ is defined by a basis $B = \{b_1, b_2, \ldots, b_k\}$ such that:

$$\mathcal{L}(B) = \left\{ \sum_{i=1}^{k} z_i b_i \mid z_i \in Z \right\} \tag{5}$$

Two central hard problems here are the following: (i) *Learning with Errors (LWE)*: Given noisy linear equations, recover the secret vector [70]. Formally, given pairs $(\boldsymbol{a}_i, b_i = \langle \boldsymbol{a}_i, \boldsymbol{s} \rangle + e_i)$, determine $\boldsymbol{s}$. The error $e_i$ is sampled from a small distribution. (ii) *Shortest Vector Problem (SVP)*: Find the shortest non-zero vector in a lattice, which is exponentially hard in general [70].

*Applications and Examples*: Lattice-based cryptography finds application in the following PQC schemes. (i) CRYSTALS-Kyber: It is a lattice-based Key Encapsulation Mechanism (KEM) and is standardized by NIST [66, 71]. It supports efficient key exchange with public keys around 1-2 KB and small ciphertexts. (ii) CRYSTALS-Dilithium: It is a lattice-based digital signature scheme with a low signature size and excellent performance [72].

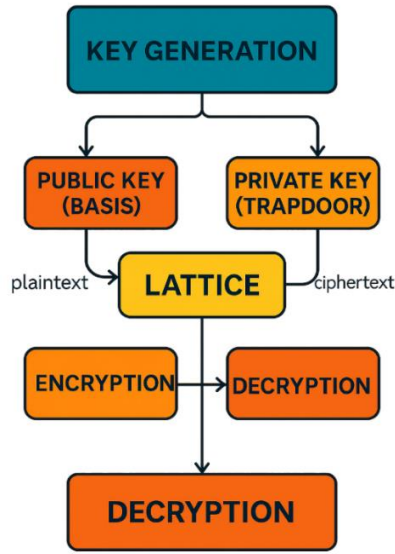

**Fig 4.7**: Operational flow of lattice-based cryptography showing key generation, encryption, and decryption.

Lattice schemes support homomorphic encryption and identity-based encryption and are resistant to side-channel attacks. Unlike RSA and ECC, which are vulnerable to Shor's algorithm (Section 4.3.2), lattice-based constructions resist both classical and quantum attacks.

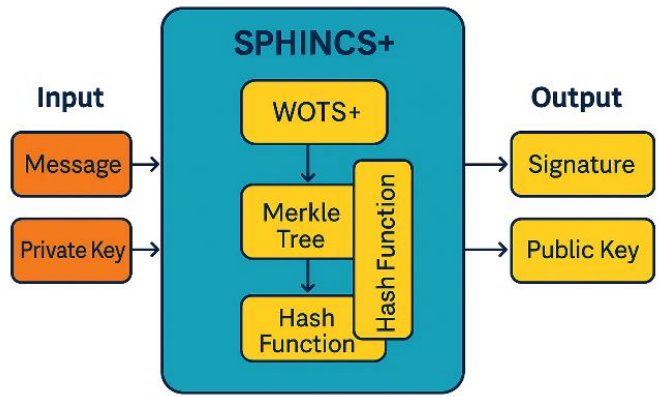

**Fig 4.8**: Architecture of the SPHINCS+ hash-based post-quantum digital signature scheme.

### 4.4.2 Hash-Based Cryptography

*Concepts and Motivation*: Hash-based cryptography derives its security from the properties of cryptographic hash functions, specifically, the difficulty of finding preimages or collisions. At the heart of hash-based signatures is the concept of one-time signatures (e.g. Lamport-Diffie) and Merkle Trees to allow for multiple secure signatures under a single public key [73]. A Merkle tree is a binary hash tree where the root represents a commitment to a large set of hash values [74]. A proof path shows that a particular hash belongs to the tree.

*Mathematical Security Assumptions*: If hash function $H$ is secure, then the scheme's security reduces to: $H(x) \neq H(x')$ for $x \neq x'$. This is also referred to as the collision-resistance property of a hash function.

*Applications and Examples*: As illustrated in Fig 4.8, SPHINCS+ is a stateless, hash-based digital signature scheme standardized by NIST [75]. It is robust, requires no state management, and is secure under minimal assumptions. Earlier hash-based schemes like XMSS required strict state tracking for security, while SPHINCS+ is stateless. This eliminates the risk of state misuse.

The main disadvantage with the hash-based schemes is their requirement of large signature sizes (usually 8-17 KB in size). The large size of signature can be a bottleneck for bandwidth-constrained or real-time IoT applications. However, their simplicity and long-term security make them ideal for firmware signing, archival documents, and satellite telemetry.

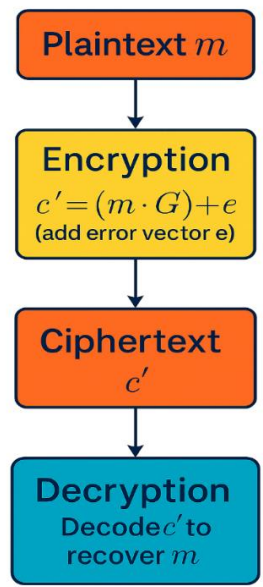

**Fig 4.9**: Schematic representation of the McEliece code-based cryptographic system.

### 4.4.3 Code-Based Cryptography

*Concept and Motivation*: Code-based cryptography relies on the difficulty of decoding general linear error-correcting codes. First introduced in 1978 by McEliece, this approach has withstood decades of cryptanalysis [76].

*Mathematical Foundation*: Let $G \in F_2^{k \times n}$ be a generator matrix of a code, where $F_2$ is the finite field with two elements. The cryptographic hardness lies in the Syndrome Decoding Problem, which is as follows: Given: (i) a generator matrix $G$ and (ii) a syndrome $s = Ge^T$, find the sparse error vector $e$. The error vector $e$ must have a small *Hamming Weight*. This problem is NP-hard and forms the basis of security for the McEliece cryptosystem [77].

*Main Scheme*: The McEliece Cryptosystem uses binary Goppa codes [78]. The public key is 1-2 MB long; the encryption and decryption both are very fast. While the large public key size is a limitation for embedded systems, it is ideal for backend systems, cloud services, and government archives, where storage is not a primary concern. Fig 4.9 exhibits the schematic representation of a code-based cryptographic system.

### 4.4.4 Multivariate Cryptography

*Concept and Motivation*: Multivariate cryptography uses the hardness of solving systems of multivariate quadratic equations over finite fields, a problem known to be NP-hard.

*Mathematical Problem*: Given a system of $m$ equations in $n$ variables: $P_j(x_1, x_2, \ldots, x_n) = y_j$ for $j = 1, 2, \ldots, m$. Each $P_j$ is a multivariate quadratic polynomial over $F_q$. The challenge is to recover the input vector $x$. This is the *Multivariate Quadratic* (MQ) *problem*, which is conjectured to be hard even for quantum computers [79].

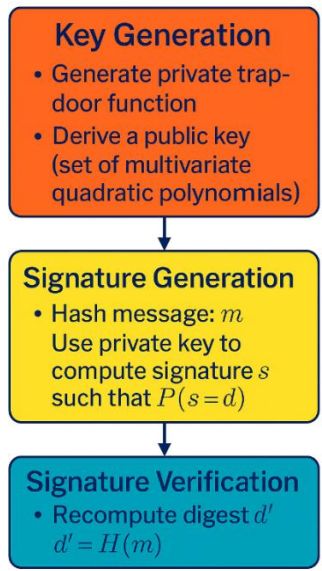

**Fig 4.10**: Structural overview of the GeMSS multivariate post-quantum signature scheme.

*Examples and Applications*: The Rainbow cryptosystem is an example of a multivariate cryptosystem [80]. It was once a finalist in the NIST PQC competition. However, it was broken in 2022 by *structural key recovery attacks* and subsequently withdrawn [81-82]. Another example of multivariate cryptosystem is GeMSS, which is a more secure variant. However, it is computationally intensive. Fig 4.10 illustrates how the GeMSS signature scheme leverages structured multivariate quadratic mappings and trapdoor functions to enable quantum-secure digital signatures. Multivariate cryptosystems are known for very fast signature generation, making them suitable for environments like smart cards and mobile authentication tokens.

A summary of leading PQC approaches, their cryptographic foundations, and current standardization status is presented in Table 4.4.

**Table 4.4:** Categories of post-quantum cryptography with hard problems, algorithms, and applications.

| PQC Type | Hard Problem | Examples | Public Key Size | Signature Size | Applications | Standardization Status |
|---|---|---|---|---|---|---|
| Lattice-Based | LWE, SVP | Kyber, Dilithium | ~1-2 KB | ~2 KB | IoT, general applications | NIST Finalist (FIPS expected 2025) |
| Hash-Based | Hash collisions | SPHINCS+ | 32 B | 8-17 KB | Firmware, archival, secure boot | NIST Alternate Candidate |
| Code-Based | Syndrome | McEliece | 1-2 MB | N/A | Backend systems, government networks | NIST Finalist |
| Multivariate | MQ problem | GeMSS, Rainbow | 100-300 KB | 10-30 KB | Mobile signatures, secure tokens | GeMSS: NIST Alternate, Rainbow: Withdrawn |

## 4.5 Quantum Cryptography in Smart Cities

The emergence of quantum cryptography marks a paradigm shift in securing digital infrastructures, particularly in highly connected urban environments. As smart cities integrate complex networks of IoT devices, cloud services, and critical infrastructure systems, ensuring secure communication and data protection becomes paramount. This section explores the conceptual foundations of smart cities and examines how quantum cryptographic mechanisms, such as QKD [83] and QRNGs [84], can be strategically embedded into their technological fabric to ensure future-proof, resilient, and privacy-preserving operations.

### 4.5.1 Smart Cities: Concepts, Components, and Challenges

A smart city refers to an urban environment that leverages digital technologies, data analytics, and interconnected infrastructures to enhance the quality of life for its residents. It optimizes the efficiency of services and infrastructure and promotes sustainable development.

At the core of a smart city is the ability to collect, analyze, and act upon large volumes of data generated by various systems. The various systems include transportation, energy, healthcare, water management, waste disposal, public safety, and so on. These systems are often embedded with sensors and actuators that are part of a broader IoT ecosystem. The collected data is transmitted through high-speed communication networks to centralized or distributed platforms, where it is processed using edge computing, cloud analytics, and artificial intelligence.

Smart cities rely on a complex technological stack comprising of the following: (i) IoT devices and sensors for real-time data acquisition, (ii) 5G and LPWAN networks [85] for high-speed, low-latency communication, (iii) cloud and edge computing for scalable and localized data processing, (iv) AI and machine learning algorithms for predictive analytics and automation, (v) cyber-physical systems (CPS) that tightly integrate computational elements with physical infrastructure.

Despite their promise, smart cities face several critical challenges. Some of them are mentioned in the following:

*Cybersecurity and Privacy*: With millions of interconnected devices continuously generating, transmitting, and processing data, the attack surface in smart cities has expanded significantly. These devices range from low-power sensors to high-end computing nodes, and may operate in unsecured or physically exposed environments. The heterogeneity of devices and protocols introduces substantial vulnerabilities, making it easier for adversaries to exploit entry points, inject malicious data, or disrupt service availability. Common cyber threats include man-in-the-middle attacks [86], spoofing [87], data exfiltration [88], ransomware [89], and denial-of-service (DoS) incidents [90] that could disable critical infrastructure such as traffic control or emergency communication systems. Moreover, centralized data aggregation platforms may become high-value targets for attackers aiming to steal or manipulate citizen data. Privacy risks are amplified by the sheer volume and granularity of data collected, when often include location, behavior, health, and biometric information.

*Interoperability*: Smart cities are typically composed of numerous subsystems deployed by different vendors. Each of them uses their own communication protocols, data formats, and proprietary software. This diversity complicates the seamless exchange of information between systems, such as traffic control networks interfacing with emergency response platforms or energy management systems communicating with public transport infrastructure. Lack of standardized interfaces can lead to siloed data, inefficient workflows, and increased integration costs.

*Scalability and Maintenance*: As urban populations grow and the demand for connected services increases smart city infrastructure must scale accordingly. This includes the ability to support more IoT devices, higher data throughput, and expanded analytics capabilities. However, scaling these systems involves several challenges, including hardware upgrading, managing power consumption, extending network coverage, and ensuring software compatibility.

*Data Governance*: The extensive data generated by smart city systems raises critical questions about ownership, consent, access control, and ethical usage. Citizens often do not have visibility or control over how their personal data collected via surveillance cameras, smart meters, or public Wi-Fi, are used, stored, or shared. Moreover, inconsistent data protection laws across jurisdictions create ambiguity in cross-border data flows. Establishing clear governance frameworks is essential to define who owns the data, how it can be used, and what rights individuals have over their information.

*Infrastructure Investment*: The journey from traditional urban infrastructure to smart city models requires substantial financial investment. Legacy systems such as analog traffic signals or standalone surveillance units need to be replaced by digital, network-enabled alternatives. This necessitates large-scale deployment of fiber-optic cables, 5G antennas, data centers, and secure edge computing nodes.

Given the scale and criticality of the services managed by smart cities, ensuring secure and resilient communication and control systems is paramount. As quantum computing advances, many classical cryptographic methods used today will become obsolete, leaving critical infrastructures vulnerable to data interception and manipulation.

### 4.5.2 Role of Quantum Cryptography in Smart Cities

In smart city environments, safeguarding data confidentiality, integrity, and authenticity is not just a technical requirement but a civic imperative. However, with the rapid advancement of quantum computing, many widely used classical cryptographic schemes, such as RSA, Diffie–Hellman, and elliptic curve cryptography (ECC), face potential obsolescence. In response to these emerging threats, quantum cryptography provides a fundamentally different approach to securing communications. The most prominent application of quantum cryptography is QKD.

*Quantum Key Distribution in Urban Communication Networks*: QKD operates by transmitting quantum states, typically photons encoded in different polarization states, through either fiber-optic cables or free-space optical channels. Protocols such as BB84 (Bennett and Brassard, 1984) [17] and E91 (based on entangled states, Ekert, 1991) [91] facilitate the establishment of symmetric keys between communicating parties. In the context of smart cities, QKD can be integrated into the communication infrastructure that connects various urban services. For example, secure high-priority communication links can be established using QKD between municipal command centers and emergency response units, such as fire departments, police stations, and hospitals. In systems like smart grids or water distribution networks, control commands sent to field devices (e.g., smart meters, valves, etc.) can be protected via QKD. This reduces the risk of command injection or unauthorized reconfiguration. Some hybrid models have been proposed where QKD is used exclusively for secure key exchange, while classical symmetric cryptographic systems (e.g., AES-256) are deployed for handling high-speed data encryption. These models address the bandwidth and latency limitations of current QKD systems. However, they maintain a quantum-safe key establishment layer. A key limitation of QKD systems is their relatively low bandwidth. The key generation rates are typically limited to a few kilobits per second (kbps) due to photon loss, channel noise, and the quantum no-cloning principle [92].

*Quantum Random Number Generators (QRNGs) for IoT Security:* Another important component of quantum cryptography in smart cities is the use of QRNGs. These random numbers are crucial for generating cryptographic keys, nonces, and initialization vectors. In contrast to pseudorandom number generators (PRNGs), QRNGs do not rely on seed values or deterministic algorithms. Hence, they provide higher entropy and greater resistance to prediction [93]. In smart cities, QRNGs can be embedded within the following entities: (i) IoT Gateways to secure edge communications, (ii) Mobile Devices for location-based citizen services, and (iii) Network Infrastructure Devices, such as routers and switches, for secure session key negotiation. Integrating QRNGs across layers of the urban infrastructure significantly raises the baseline of the cryptographic strength associated with the key generation algorithms.

*Metropolitan QKD Networks and Trusted Node Models*: For large-scale deployment, smart cities can develop Metropolitan Quantum Networks (MQNs) [94]. These are dedicated fiber-optic infrastructures that connect trusted nodes at key urban locations. These trusted nodes act as intermediate points that measure and regenerate quantum signals while preserving the integrity of the key exchange. Cities such as Beijing, Tokyo, and Vienna have already piloted MQNs, linking banks, government facilities, and data centers using QKD-protected channels [18].

In the trusted-node-based QKD, quantum keys are generated between adjacent nodes and relayed hop-by-hop across the network. Each intermediate node must be physically trusted, as it has access to key material during regeneration. In contrast, entanglement-based key exchange, currently a subject of active research, enables direct end-to-end quantum key generation between distant users without requiring trust in intermediate nodes. This is achieved by distributing entangled photon pairs and using quantum repeaters to maintain correlations over long distances, though practical deployment of such systems is not yet mature [95-96].

Since a compromised node can potentially leak keys, the trusted node model introduces physical security. Hence, it remains the most practical form of QKD deployment at metropolitan scales until fully end-to-end quantum repeaters become commercially viable.

*Integration with Urban Security Frameworks*: Quantum cryptographic mechanisms can be embedded within broader security frameworks in a mart city through various systems such as: (i) *Security Information and Event Management (SIEM) systems*, (ii) *Public Key Infrastructure (PKI),* (iii) *Blockchain-based Identity Systems*, etc. In SIEM, quantum-resilient encryption can protect logs and alerts generated by smart city applications [97-98]. In PKI, certificate and key distribution methods can be upgraded using post-quantum secure digital signatures, with QKD used for the transport of symmetric session keys [99-101]. In decentralized identity (DID) schemes, quantum-enhanced blockchain frameworks can incorporate QRNG-generated keys and hash-based or lattice-based signatures for identity validation [102-103].

### 4.5.3 Reference Architectures for Quantum-Secured Smart Cities

In the context of smart cities, a robust and scalable architecture is essential for integrating quantum cryptography into existing digital infrastructure. Quantum-secured smart cities require multi-layered security architectures incorporating both classical and quantum components. One of the foundational models proposed is the Quantum-Secured Smart City Architecture (QSSCA) [104]. As exhibited in Fig 4.11, QSSCA consists of four principal layers: (i) Device Layer, (ii) Network Layer, (iii) Platform Layer, and (iv) Application Layer. The features of these layers are discussed briefly in the following.

*Device Layer*: This layer includes a diverse set of entities such as sensors, actuators, smart meters, wearables, surveillance equipment, and citizen devices. Devices in this layer are often constrained in terms of computational resources and power. This makes them vulnerable to cyberattacks. Embedding QRNGs in these devices enhances their security.

*Network Layer*: The network layer provides the physical and logical pathways for data exchange. In a quantum-secured city, this includes the deployment of QKD channels over fiber-optic networks and potentially free-space optical links. Quantum routers, switches, and trusted nodes are used to distribute symmetric keys across city sectors. These nodes serve as intermediary points and perform real-time authentication and key agreement.

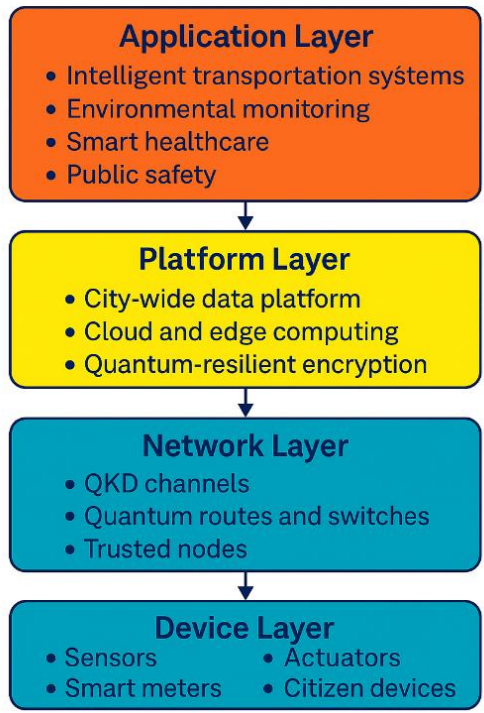

**Fig 4.11**: Layered architecture of a quantum-secured smart city (QSSCA).

*Platform Layer*: This layer includes city-wide data platforms, cloud services, and edge computing nodes responsible for analytics, decision-making, and orchestration. Quantum-resilient encryption protocols and PQC primitives are embedded in platform APIs to ensure that sensitive data at rest or in motion is protected against current and future quantum threats.

*Application Layer*: This layer hosts the operational services of the smart city, including *intelligent transportation systems* (ITS), environmental monitoring, smart healthcare, public safety, and emergency services. These applications interface with both citizens and infrastructure, making them critical points of vulnerability. Integration of quantum-aware APIs ensures that these applications can perform secure computing and communication using QKD and PQC algorithms.

*Security Management Modules and Operational Control*: Cross-cutting all the above four layers is a security management module that includes a SIEM system enhanced with quantum-assisted threat detection algorithms. These systems are responsible for aggregating logs, detecting anomalies, and initiating automated incident response using policies defined in Quantum-Integrated Security Orchestration, Automation, and Response (Q-SOAR) [105].

A model deployment could involve a city's traffic management authority using QKD to secure data links between control centers and roadside units, a smart healthcare system employing QRNGs for patient data security, and government services using PQC digital signatures to issue tamper-proof documents.

### 4.5.4 Hybrid Quantum-Classical Security Models

While the goal of achieving end-to-end quantum-secured communication in smart cities remains on the horizon, hybrid quantum-classical security models offer an immediate and realistic approach to enhancing urban cybersecurity. These models combine the robust, well-understood classical cryptographic mechanisms with the forward-looking potential of quantum cryptography. As shown in Fig 4.12, A practical hybrid model in a smart city includes the following layers: (i) quantum layer, (ii) classical layer, (iii) integration layer, and (iv) application layer.

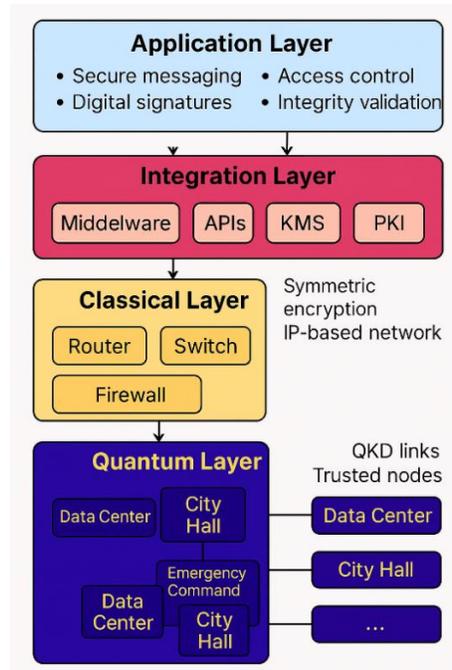

**Fig 4.12**: Hybrid framework combining classical and quantum secure communication layers.

The quantum layer implements QKD links using point-to-point fiber optic infrastructure or free-space optical channels. Trusted nodes are deployed at strategic locations such as data centers, city halls, and emergency command units. The classical layer comprises the conventional IP-based communication network, routers, switches, and firewalls that transmit encrypted payloads using keys negotiated by the quantum layer. The integration layer includes middleware and APIs that interface QKD hardware with legacy key management systems (KMS) and public key infrastructure (PKI). This ensures smooth interoperation between quantum-generated keys and classical security protocols. Finally, the application layer ensures secure messaging, access control, digital signatures, and integrity validation across services like traffic control, public transportation, healthcare utilities, and surveillance systems.

The evolving landscape of security models across classical, hybrid, and fully quantum systems is summarized in Table 4.5, highlighting their strengths, limitations, and implementation readiness.

**Table 4.5:** Comparison of classical, hybrid, and quantum security models.

| Security Model | Key Exchange | Encryption Algorithm | Quantum Resistance | Infrastructure Requirement | Adoption Readiness |
|---|---|---|---|---|---|
| Classical | RSA, ECDH | AES, RSA, ECC | Vulnerable to Shor's / Grover's | Existing infrastructure only | High (Current standard) |
| Hybrid Quantum-Classical | QKD + Classical KMS | AES-256, ChaCha20 | Partially quantum-resistant | QKD links + existing classical net | Medium (Incremental upgrade) |
| Full Quantum | Entanglement/ QKD + QRNG | Quantum-secure primitives | Provably quantum-secure | Quantum repeaters, quantum memory | Low (Still under development) |

*Real-Word Deployment Scenarios*: Several real-world deployments have demonstrated the feasibility of hybrid quantum-classical models in urban settings. *Beijing-Shanghai Backbone Network* is a 2000-km network that connects major cities in China using QKD-secured key distribution combined with classical communication protocols [106]. *Vienna Quantum Network* provides QKD between nodes and integrates classical cryptographic schemes to secure governmental communications [18]. *SwissQuantum Project* validated hybrid encryption of financial data between banks using QKD and AES [107].

*Path Forward*: City planners and security architects should prioritize deploying QKD links between critical urban nodes, such as data centers and public safety agencies. In practical deployment hybrid architectures will provide a transitional mechanism, blending classical and quantum cryptographic elements. A common configuration is a quantum-secure key distribution using QKD to negotiate session keys, while classical symmetric key encryption using AES-256 to handle high-throughput data flows. Urban deployment models may use Quantum Trusted Nodes (QTNs), which act as secure exchange points within the city. These nodes enable point-to-point QKD sessions and are integrated with conventional Public Key Infrastructure (PKI) for managing digital certificates. The PQC algorithms, such as CRYSTLS-Dilithium and SPHINCS+, described in detail in Section 4.4, offer digital signature schemes that are resistant to quantum attacks and integrate naturally into hybrid architecture alongside QKD-secured key exchange.

### 4.5.5 Quantum Cryptography Applications in Smart City Domains

Quantum cryptography holds transformative potential across a wide array of smart city domains. Smart Transportation, Smart Energy, e-Governance and Public Services, Smart Healthcare Systems, Surveillance and Law Enforcement, etc. are some very important applications of quantum cryptography in smart cities. These applications are discussed briefly in the following.

*Smart Transportation*: Intelligent transportation systems (ITS), autonomous vehicles, and traffic coordination platforms rely heavily on low-latency, high-integrity communication between vehicles (V2V), infrastructure (V2I), and control centers (V2C) [108]. QKD can be employed to secure these channels by establishing tamper-evident session keys for critical message exchange. For instance, dynamic traffic rerouting signals or hazard alerts between smart traffic lights and autonomous cars must be shielded from spoofing and interception. QRNG-enhanced tokens can provide added entropy in vehicular authentication processes.

*Smart Energy*: Modern energy distribution systems, integrate smart grids, distributed energy resources (DERs), and demand-response mechanisms. These systems depend on resilient control signals and privacy-preserving metering. Quantum cryptography can ensure the secure communication of load-balancing commands, voltage regulation data, and fault notifications between substations and control centers. By incorporating QKD into SCADA (Supervisory Control and Data Acquisition) systems, utilities can prevent man-in-the-middle attacks on real-time operational data [18].

*e-Governance and Public Services*: Public services increasingly involve electronic transactions, identity validation, and document authentication through digital portals. Quantum-secure communication ensures that *personally identifiable information* (PII) and legal credentials (e.g., birth certificates, property titles, municipal permits) are immune to eavesdropping or tampering. QKD-enabled secure tunnels can connect government buildings and data repositories, while lattice-based or hash-based digital signatures can replace RSA-based certificates in government-issued digital IDs.

*Smart Healthcare Systems*: Electronic health records (EHRs), wearable monitoring systems, and remote diagnostics require strict protection of patient data. In smart cities, medical IoT devices often communicate over shared wireless networks, making them vulnerable to interception. QKD can be used to secure links between hospitals, insurance providers, and data analytics platforms. Secure multi-party computation protocols enhanced with PQC primitives can also allow for federated learning on medical data without exposing raw datasets [109]. To further enhance privacy, federated learning techniques can be combined with post-quantum secure communication channels.

*Surveillance and Law Enforcement*: Surveillance systems generate high-volume video and audio feeds transmitted over urban networks. Quantum-enhanced encryption techniques help ensure that this sensitive data, used for forensic analysis, crowd monitoring, and emergency detection, remains intact and unaltered. QKD can protect data streams from being intercepted or tampered with. Further, smart surveillance systems must comply with evolving privacy regulations such as the GDPR [110]. Techniques like privacy-preserving video analytics, including on-device processing and anonymized facial recognition, can be combined with quantum-secure encryption to ensure compliance while retaining forensic utility.

The implementation of quantum cryptography in these domains is not merely theoretical. Pilot deployments and testbeds in global smart cities (e.g., Toronto, Vienna, Shanghai) have demonstrated the feasibility of integrating QKD and PQC with existing infrastructure [99].

### 4.5.6 Standards and Protocols for Urban Quantum Security

A consistent, secure, and interoperable deployment of quantum cryptography in smart cities necessitates strict alignment with emerging international standards. Some of the important standardization bodies and their activities on quantum cryptography are briefly discussed in the following.

*European Telecommunication Standards Institute Industry Specification Group on QKD (ETSI ISG-QKD)*: The ETSI ISG-QKD is one of the most active international bodies working on the standardization of QKD networks [111]. Its specifications cover system architecture, security requirements, key lifecycle management, authentication procedures, and interoperability of QKD components. It has defined a reference architecture for integrating QKD systems into classical networks, including secure key relay, quantum channel management, and control interfaces.

*NIST Post-Quantum Cryptography (PQC) Standardization*: NIST of the USA leads the development of cryptographic algorithms that are resistant to attacks by quantum computers. In 2022, NIST selected the following four primary candidates for standardization: (i) CRYSTALS-Kyber [66], (ii) CRYSTALS-Dilithium [66,72], (iii) FALCON [112], and (iv) SPHNICS+ [113]. CRYSTALS-Kyber is a lattice-based key encapsulation mechanism built upon the hardness of the *Module Learning with Errors* (MLWE) problem [69]. It is designed for key exchange and public-key encryption with small ciphertexts and high performance, making it suitable for resource-constrained IoT devices. Its resistance to side-channel attacks makes it highly applicable in smart city sensors and mobile devices. CRYSTALS-Dilithium is a lattice-based digital signature algorithm based on the Module-SIS (MSIS) [72] and MLWE [69] problems. It supports authentication in secure boot processes, digital document signing, and identify validation within municipal services. Fast-Fourier Lattice-based Compact Signatures over NTRU (FALCON) is a signature scheme based on NTRU lattice problem [114]. It offers shorter signature sizes and faster verification. This makes it ideal for bandwidth-sensitive applications such as firmware signing and certificate validation in constrained devices and networks. However, its use of floating-point arithmetic introduces implementation challenges, including potential side-channel vulnerabilities, which demand careful attention during integration. SPHINCS+ is a stateless hash-based signature scheme built on Merkle Tree constructions [74]. Although its signature sizes are longer which make the verification slower, it is highly robust and suitable for long-term archival, legal documents, or use cases that prioritize durability over performance.

*ITU-T Recommendation for QKD*: The International Telecommunication Union (ITU), through Study Group 13 (SG13) is developing recommendations on the integration of QKD into telecom and urban digital ecosystems [115]. The key documents contributed by ITU-T in quantum cryptography include (i) Y.3800 series, which defines the functional architecture, components, and interfaces of QKD networks, and (ii) recommendations on trusted repeater security, quantum-classical key relay protocols, and end-to-end encryption workflows.

*Interoperability Frameworks and Certification*: The OpenQKD project [116-117] in Europe and the Quantum Communications Hub [118] in the UK aim to develop interoperable multi-vendor quantum networks with shared APIs. Certification labs are being established to validate the conformance of QKD hardware, QRNGs, and PQC software to published standards. The National Cybersecurity Center of Excellence (NCCoE) in the U.S. publishes guidelines for migration to post-quantum algorithms in governments and enterprise networks [119].

Fig 4.13 summarizes the contributions of various international standardization bodies toward establishing secure quantum communication frameworks for smart cities.

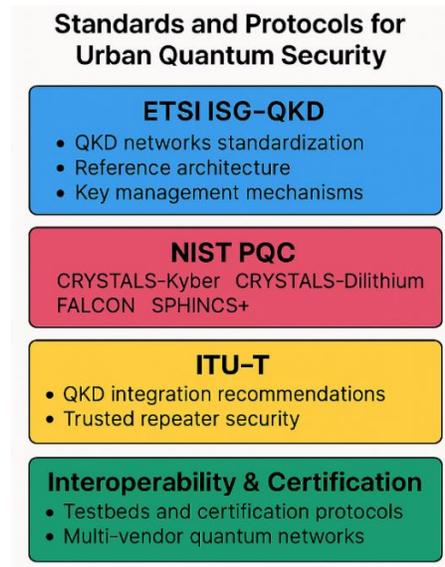

**Fig 4.13**: Overview of global standardization efforts in urban quantum security.

A comparative view of quantum technologies and their practical applications across various domains is presented in Table 4.6, outlining their key advantages and challenges.

**Table 4.6:** Quantum-enabled tools and their applications across critical sectors.

| Quantum Tool | Application Layer | Key Benefits | Deployment Considerations |
|---|---|---|---|
| **QKD** | Transportation, Governance, Energy, Healthcare | - Information-theoretic key exchange<br>- Tamper detection<br>- Resilience to APT | - Requires dedicated fiber or free-space optical links<br>- Trusted nodes introduce physical security risks |
| **QRNG** | IoT Devices, Mobile Services, Surveillance | - High-entropy, truly random key material<br>- Improves cryptographic strength in edge devices | - Integration into low-power hardware<br>- Must comply with entropy validation standards (e.g., NIST SP 800-90) |
| **PQC (e.g., Kyber, Dilithium, SPHINCS+)** | e-Governance, Public Services, Data Centers | - Resistance to quantum algorithms<br>- Works on classical hardware<br>- Suitable for digital signatures and authentication | - Larger key/signature sizes<br>- Transition needs cryptographic agility<br>- Compliance with emerging NIST standards |
| **Hybrid QKD + AES/PQC** | Cross-domain Secure Messaging | - Combines QKD's secure key exchange with AES/PQC's scalability<br>- Enables layered defense | - Middleware integration complexity<br>- Legacy compatibility must be ensure |
| **Quantum-Secure Blockchain (e.g., with PQC + QRNG)** | Identity Management, Voting, Smart Contracts | - Immutable, verifiable ledgers with quantum-resilient keys<br>- Robust against future attacks | - Cryptographic redesign of consensus mechanisms<br>- Storage overhead for PQC-enhanced transactions |

## 4.6 Security and Privacy Management in the IoT using Quantum Cryptography

The motivation for securing the IoT in the post-quantum era stems from the convergence of two timelines: the projected advancement of quantum computing and the deployment life cycles of IoT devices, many of which remain active in the field for a decade or more. If secure communication channels are compromised by quantum adversaries in the future, the implications for data confidentiality, device integrity, and public trust could be severe. Hence, it becomes imperative to explore and implement quantum-safe alternatives, such as PQC [14], QKD [83], and QRNGs [84].

This section outlines the limitations of classical cryptographic primitives in IoT environments and sets the stage for a detailed exploration of quantum-enhanced solutions that aim to protect the integrity and privacy of IoT ecosystems well into the quantum future.

### 4.6.1 Privacy Challenges in IoT Environments

The proliferation of IoT technologies across domains such as healthcare, smart homes, urban infrastructure, and industrial automation has led to an unprecedented collection of personal, environmental, and operational data. IoT devices continuously monitor and transmit granular information, ranging from biometric readings and geolocation data to behavioral patterns and machine diagnostics. This extensive data collection creates vast digital footprints that are deeply sensitive in nature and susceptible to exploitation.

*Illustrative Breaches in IoT Privacy*: Several high-profile cases highlight the practical implications of IoT privacy vulnerabilities. The *Mirai Botnet Attack* exploited weak default passwords in IoT devices like CCTV cameras and home routers, turning them into a botnet that disrupted global internet traffic via large-scale DDoS attacks [120-121]. *Ring Camera Intrusion* involved unauthorized access to home security cameras, where attackers gained control over camera feeds and engaged in direct verbal harassment, revealing deep vulnerabilities in device authentication and cloud linkage [122]. In Philips Hue Smart Lighting Systems, researchers demonstrated an exploit in which attackers could remotely infect devices and control entire networks by using a drone outside buildings [123-124]. These real-world incidents emphasize the urgent need for robust, privacy-respecting design in the increasingly ubiquitous IoT landscape.

*Nature and Sensitivity of Data*: IoT ecosystems usually collect data that goes beyond standard identifiers. Smartwatches monitor heart rate variability and physical activity. Smart home sensors track occupancy, light usage, and voice commands. Industrial IoT nodes log machine telemetry and predictive maintenance metrics. These data are essential for automation and personalization. However, they can also reveal finer details about individuals' health, habits, movements, and even emotional states. When aggregated over time, these data streams enable highly granular behavioral profiling, raising critical concerns about user privacy, surveillance, and consent [125-127].

*Risks of Eavesdropping, Inference Attacks, and Profiling*: Due to the wireless nature and distributed deployment of many IoT devices, communication channels are particularly vulnerable to eavesdropping. Adversaries can intercept unencrypted or weakly encrypted transmissions to gain unauthorized insights. Beyond direct data interception, attackers can perform inference attacks, exploiting correlations within benign data streams to uncover sensitive attributes. For instance, energy consumption patterns can reveal occupancy or sleep schedules, while traffic from a smart medical device may expose chronic health conditions [109,125].

*Data Lifecycle Vulnerabilities*: IoT data pass through several stages, from collection at edge devices, to transmission across wireless or wired networks, to storage in local or cloud servers, and finally to processing for analytics or automation. Each of these stages presents unique data privacy challenges. In the collection phase, the sensors often operate without user awareness or explicit consent [126-128]. This leads to the passive accumulation of sensitive data. During the transmission phase, many devices lack secure communication protocols. This leads to transmitted data prone to interception or modification. In the data storage phase, cloud platform storing IoT data become high-value targets for attackers. Poor access controls or data leaks can result in large-scale breaches. Finally, data processing done by AI systems or data shared across third parties can be done in opaque ways, violating user expectations and privacy agreements.

**Table 4.7:** Classification of privacy risks in the IoT by system layer and attack vector.

| Source Layer | Example Components | Types of Privacy Risk | Typical Attacks/Breaches |
|---|---|---|---|
| **Device Layer** | Sensors, wearables, smart cameras | Identity exposure, location tracking, eavesdropping | Device fingerprinting, metadata leakage |
| **Communication Layer** | Wi-fi, Bluetooth, Zigbee, 5G modules | Interception, spoofing, man-in-the-middle (MITM) | Unencrypted data transmission, protocol attacks |
| **Aggregation Layer** | IoT hubs, gateways, edge servers | Inference attacks, unauthorized profiling | Traffic analysis, pattern learning |
| **Storage Layer** | Cloud servers, local memory units | Unauthorized access, retention beyond consent | Misconfigured cloud storage, unauthorized backups |
| **Application Layer** | Smart apps, dashboards, analytics engines | Over-collection, consent violation | Data misuse, shadow profiling |

These vulnerabilities are exacerbated by the resource-constrained nature of IoT hardware, which may lack the computational capabilities to implement strong encryption, intrusion detection, or user authentication mechanisms. Moreover, many devices lack long-term maintenance support, meaning security patches and privacy-preserving upgrades are rarely applied, leaving data exposed for years. Table 4.7 exhibits various privacy risks in the IoT.

The following sections explore how quantum-enhanced technologies, ranging from PQC to QKD and QRNG, can be strategically deployed to protect user privacy in IoT systems.

### 4.6.2 Quantum Threats to Classical IoT Security

The anticipated rise of large-scale quantum computers presents a critical and inevitable threat to the security mechanisms in IoT infrastructures. Classical cryptographic algorithms, especially those based on number-theoretic assumptions, are vulnerable to quantum attacks [129]. As IoT systems are increasingly deployed in vital services such as energy management, urban transportation, healthcare, and national defense, securing them against quantum adversaries is no longer optional but mandatory.

Two cornerstone quantum algorithms pose the most significant threats to the current encryption mechanisms. These are: (i) Shor's Algorithm and (ii) Grover's Algorithm.

*Shor's Algorithm*: Introduced by Peter Shor in 1994, this quantum algorithm allows polynomial-time factorization of large integers and efficient computation of discrete logarithms [60]. Since the classical public-key systems such as RSA [34], Diffie-Hellman [35], and Elliptic Curve Cryptography (ECC) [36-37] are based on the assumptions of infeasibility of polynomial-time factorization of large primes and efficient computation of the discrete logarithms, the security of these algorithms is under considerable threat with the advent of Shor's algorithm. Once quantum computers with sufficient qubit stability and fault-tolerant error correction are realized, Shor's algorithm will be able to break 2048-bit RSA or 256-bit ECC keys in seconds, rendering the entire public-key infrastructure vulnerable. In the context of the IoT, this is especially alarming given that many embedded systems implement these algorithms on secure bootloaders, firmware signing, and over-the-air update verification process.

*Grover's Algorithm*: Proposed in 1996 by Lov Grover, this quantum algorithm provides a quadratic speedup for unstructured search problems [12]. In cryptographic terms, Grover's algorithm significantly affects symmetric-key encryption schemes by reducing the brute-force effort required to find a secret key. Specifically, if a symmetric cipher like AES-128 requires $2^{128}$ operations to exhaustively search its key space using classical approaches, Grover's algorithm reduces this complexity to roughly $2^{64}$ operations on a quantum computer. While this is not a complete break of the algorithm, it effectively halves the key size, making brute-force attacks far easier in a quantum world. As a mitigation strategy, doubling the key length is recommended. Hence, moving from AES-128 to AES-256 is required to restore the desired security margin [14]. However, in the context of the IoT, this approach encounters practical limitations. Many IoT devices are resource-constrained, which makes it difficult or even infeasible to implement heavy cryptographic workloads such as AES-256 on them. Moreover, latency-sensitive IoT applications, such as those used in autonomous vehicles or medical telemetry, may suffer from increased computational overhead. Hence, the deployment of symmetric cryptographic algorithms like AES remains a significant challenge in IoT landscape.

Many communication protocols in use across IoT networks such as MQTT for telemetry data [29], CoAP for constrained web-based resources [30], and DTLS for transport-layer security [31], depend heavily on RSA or ECC for key negotiation and authentication. As such, their current implementations are directly exposed to quantum vulnerabilities.

An essential cryptographic property to consider in this transition is forward secrecy. In classical cryptography, forward secrecy ensures that the compromise of long-term keys does not expose past session data. This is typically achieved in ephemeral Diffie-Hellman exchanges. In the post-quantum landscape, lattice-based ephemeral *key encapsulation mechanisms* (KEMs), such as those derived from CRYSTALS-Kyber [66, 71], are being developed to provide equivalent guarantees. These enable encrypted sessions to remain secure even in the face of future key disclosure, a principle that becomes especially critical in the post-quantum era.

*The "Harvest Now, Decrypt Later" Paradigm*: One of the most dangerous and often underestimated strategies enabled by the development of quantum computing is the "Harvest Now, Decrypt Later" (HNDL) paradigm [13]. This approach involves adversaries intercepting and storing encrypted communications today, even if they are currently indecipherable, on the assumption that future quantum computers will possess the computational capacity to decrypt them with ease. Unlike conventional attacks, which aim for real-time interception and decryption, HNDL represents a long-term strategic threat with significant implications for data privacy.

The HNDL model is particularly threatening for IoT systems due to: (i) long data retention lifespan, (ii) prevalence of quantum-susceptible algorithms, (iii) legal and regulatory risks, (iv) impact on national security and critical infrastructure, and (v) cascading effects in the digital supply chain.

*Quantum-vulnerable Algorithms*: Public-key algorithms like RSA-2048 and ECC-256 are still widely used in IoT protocols for key exchange, firmware signing, and device authentication. These algorithms are insecure against Shor's algorithm. Encrypted sensor logs, device logs, authentication tokens, and session records, even if encrypted securely under today's assumptions, will become accessible post-quantum if not resecured or re-encrypted with quantum-resilient methods.

**Table 4.8:** Impact of classical and quantum threats on critical privacy dimensions in secure communication systems.

| Privacy Dimension | Classical Threats | Quantum Threats |
|---|---|---|
| Confidentiality | Brute-force decryption, side-channel attacks | Shor's algorithm breaks RSA/ECC; Grover's reduces AES-128 or 64-bit effort |
| Anonymity | Device fingerprinting, metadata inference | Stored encrypted traffic deanonymized post-quantum |
| Traceability | Session correlation, IP/MAC reuse | Post-quantum attacks on pseudonymous logs and surveillance metadata |
| Forward Secrecy | Relies on ephemeral Diffie-Hellman | PQC-based key exchange (e.g., Kyber) needed to ensure resilience |
| Integrity | Weak hashes (e.g., SHA-1 collisions), spoofed updates | Grover's algorithm reduces strength of hash-based authentication |
| Authentication | Credential replay, PKI spoofing | Shor breaks digital signatures (RSA/ECC); need PQ-safe alternatives |

*Legal and Regulatory Risks*: Regulatory frameworks such as the General Data Protection Regulation (GDPR) [110], the Health Insurance Portability and Accountability Act (HIPAA) [130], and emerging international data protection laws [131] emphasize the importance of long-term confidentiality and integrity. Organizations that experience data breaches in the future, due to HNDL-enabled quantum decryption, may be found non-compliant with such standards, even if their encryption practices were deemed adequate at the time of collection. This introduces liability risks, potential fines, and reputational damage.

*National Security and Critical Infrastructure*: Many government and defense IoT networks rely on encrypted telemetry, sensor data, and surveillance channels that adversarial state actors may silently harvest . If decrypted in the future, such data could reveal operational patterns, infrastructure weaknesses, or personnel movements. This may jeopardize strategic security. The threat extends to critical infrastructure operators, such as power grids, water supply networks, and transportation systems [132-133].

The implications of classical and emerging quantum threats on core privacy properties are summarized in Table 4.8, highlighting the need for quantum-resilient cryptographic measures.

Given the above vulnerabilities of the current security protocols, the adoption of forward-secure encryption mechanisms, which provide resilience against future decryption even if the long-term keys are compromised, is imperative. Techniques such as post-quantum cryptographic algorithms (e.g., lattice-based KEMs), ephemeral key exchange protocols, and quantum-safe digital signatures should be prioritized.

Recent studies, including those by IBM and the Quantum Economic Development Consortium (QED-C), suggest that *cryptographically relevant quantum computers* (CRQCs) are expected to become viable within the next 10–15 years [129, 134-135]. In that event, factoring a 2048-bit RSA key may require around 4000 logical qubits, supported by hundreds of thousands to millions of physical qubits with sufficient error correction layers. This timeframe highlights the urgency for cryptographic transitions in IoT infrastructure.

The existing IoT ecosystems rely on insecure classical algorithms across various protocols as depicted in Table 4.9. Mitigation strategies are needed as a long-term security against such threats.

**Table 4.9:** Quantum vulnerabilities in common IoT communication protocols and corresponding mitigation strategies.

| Protocol | Algorithm Used | Quantum Risk | Mitigation Strategy |
|---|---|---|---|
| MQTT | RSA/ECC | Broken by Shor | PQC KEM or Hybrid TLS |
| DTLS | ECC | Broken by Shor | Post-Quantum Handshake |
| CoAP | ECC/RSA | Broken by Shor | PQC-Capable Cipher |
| TLS1.3 | ECC | Key Exchange Vulnerable | Hybrid Key Agreement |

Given these vulnerabilities, immediate steps should be taken to integrate post-quantum security standards into these protocol stacks. Several hybrid cryptographic models, which combine classical and post-quantum primitives, are being developed and tested in real-world environments to enable secure communication without compromising existing interoperability.

The next section explores in detail the emerging post-quantum cryptographic primitives that are best suited for deployment in constrained IoT environments, including their performance characteristics, implementation challenges, and current standardization progress.

### 4.6.3 Post-Quantum Cryptographic Approaches for the IoT

The looming threat of quantum-enabled adversaries has catalyzed a global research initiative to identify cryptographic systems capable of withstanding quantum attacks. These algorithms, collectively referred to as post-quantum cryptographic (PQC) primitives, are built on mathematical problems believed to remain hard even for quantum computers. For IoT systems, these schemes must balance strong security guarantees with the computational, memory, and energy constraints intrinsic to embedded platforms. This section explores leading families of PQC schemes and their implications for secure IoT deployment.

*Lattice-Based Cryptography*: Lattice-based cryptographic schemes are widely considered the most promising candidates for post-quantum security due to their strong theoretical foundation, efficiency, and versatility. CRYSTALS-Kyber is a lattice-based Key Encapsulation Mechanism (KEM), has been selected by NIST as the standard for quantum-resistant public key encryption [66, 71]. CRYSTALS-Dilithium is a companion signature scheme based on the same lattice assumptions, providing high-speed digital signatures with small signatures and public keys [72].

*Hash-Based Signatures*: Hash-based signatures derive their security solely from the security of the underlying hash function, making them conceptually simple and highly trustworthy. Among them, SPHINCS+ stands out as a stateless, hash-based digital signature scheme suitable for post-quantum environments [75].

*Code-Based Cryptography*: The McEliece cryptosystem [76-77], proposed in 1978, is based on the difficulty of decoding a general linear code, particularly binary Goppa codes [78]. It has withstood decades of cryptanalytic scrutiny and is considered one of the most mature post-quantum encryption schemes.

*Multivariate Quadratic Schemes*: These cryptographic schemes leverage the mathematical hardness of solving multivariate quadratic polynomial systems over finite fields [79]. This problem is NP-hard and is believed to resist quantum attacks. Signature schemes like Rainbow and GeMSS are representatives of this class [79, 136]. These schemes generally exhibit high-speed signature generation and relatively small private keys. However, they suffer from either large public keys, as in Rainbow, or large signature sizes, as in GeMSS.

*Isogeny-Based Cryptography*: Isogeny-based schemes rely on the hardness of finding isogenies between supersingular elliptic curves. This problem is thought to be quantum resistant. The most prominent example, SIKE (Supersingular Isogeny Key Encapsulation) had extremely small key and ciphertext sizes, often less than 500 bytes [137]. This made it seemingly ideal for IoT devices with minimal memory and transmission budgets.

*Performance Considerations and Implementation Trade-Offs*: When selecting a post-quantum cryptographic scheme for IoT applications, several trade-offs must be addressed. This is because there is no single solution that universally optimizes security, performance, and resource utilization. These considerations are especially critical in IoT context, where devices are often constrained by computational power, energy availability, and memory size. Many post-quantum schemes require significantly larger public keys and signatures compared to traditional RSA or ECC-based methods. For instance, SPHINCS+ can produce signatures in the range of tens of kilobytes, which is impractical for many IoT devices [75]. On the other hand, lattice-based schemes like CRYSTALS-Kyber [66, 71] and Dilithium [72] offer more compact key and signature sizes, generally under 2 KB, making them better suited for constrained environments. Nevertheless, even small increases in cryptographic overhead can impact system design.

*Computation Overhead*: Some PQC schemes introduce a higher computational load during encryption, decryption, signing, or verification. This can lead to slower execution times and increased energy consumption. To address this, developers must explore hardware acceleration (e.g., cryptographic co-processors or FPGA-based designs) or lightweight software optimizations tailored to specific architectures.

*Bandwidth Usage*: Communication overhead becomes a concern when key exchange protocols or digital signatures produce large ciphertexts or message authentication tags. This is especially relevant in wireless IoT deployments, where network efficiency translates directly into battery life and latency. Schemes with larger key material can lead to packet fragmentation, retransmissions, and greater vulnerability to packet loss. These may adversely affect the quality of service in time-sensitive applications such as industrial automation or autonomous systems.

**Table 4.10:** Evaluation of post-quantum cryptographic algorithms for IoT suitability.

| Algorithm | PQC Type | Key/Signature Size | Pros | Challenges | Suitability |
|---|---|---|---|---|---|
| CRYSTALS-Kyber | Lattice-based (KEM) | ~1 - 2 KB | Compact, fast, side-channel resistant | Moderate RAM usage in encaps/decaps | High |
| CRYSTALS-Dilithium | Lattice-based (Signature) | ~2 – 3 KB signature, ~1 – 2 KB key | Stateless, efficient, deterministic signing | May strain RAM in tiny microcontrollers | High |
| SPHINCS+ | Hash-based (Signature) | 8 – 17 KB signature | Stateless, strong security, simple assumptions | Large signatures, slow verification | Low - Moderate |
| McEliece | Code-based (KEM) | 1 – 2 MB public key | Fast encryption, robust legacy | Impractically large keys for IoT nodes | Low |
| FALCON | Lattice-based (Signature) | ~666 B signature, ~1 – 2 KB key | Compact signatures, efficient verification | Complex FP arithmetic, side-channel concerns | Moderate |
| GeMSS | Multivariate (Signature) | Multivariate (Signature) | Fast signing | High memory and bandwidth footprint | Low-Moderate |

*Implementation Security*: In addition to mathematical robustness, implementation security remains paramount in the IoT. Many devices operate in hostile or physically accessible environments. This makes them vulnerable to side-channel attacks such as timing analysis, power monitoring, and electromagnetic leakage. Schemes like CRYSTALS-Dilithium [72] and Kyber [66, 71] with constant-time operations are resistant to known side-channel vectors. Furthermore, secure firmware development practices, fault injection countermeasures, and runtime cryptographic audits should be incorporated into the PQC integration process.

Careful benchmarking and context-aware evaluation are necessary before selecting a PQC scheme for deployment. Table 4.10 presents a comparative analysis of leading PQC algorithms based on type, key size, advantages, challenges, and suitability for IoT applications.

### 4.6.4 Integration of Quantum Randomness in IoT Privacy

One of the foundational pillars of secure communication systems is entropy. Entropy of a system reflects its degree of randomness. In the post-quantum era, the need for high-quality, unpredictable randomness becomes even more pronounced due to the increased reliance on cryptographic mechanisms that are resistant to quantum attacks. This section examines the integration of QRNG into IoT infrastructure and how they enhance privacy, improve cryptographic strength, and provide entropy sources for constrained devices.

#### 4.6.4.1 Role of Quantum Random Number Generators (QRNGs)

Random number generation forms the backbone of cryptographic systems, as the unpredictability and uniformity of random values directly influence the strength of encryption, authentication, and digital signatures. Traditional random number generators (RNGs), particularly those used in constrained IoT devices, are typically pseudo-random, as they rely on algorithms seeded by system time, electrical noise, or internal counters. While pseudo-random number generators are effective in general-purpose computing, such sources are often insufficient in IoT contexts, in which limited entropy, repeated startup states, or power-cycling can cause RNGs to produce weak or even predictable outputs [93, 138].

QRNGs overcome these limitations by leveraging the intrinsic uncertainty of quantum mechanics [139]. Specifically, QRNGs extract entropy from quantum phenomena such as photon arrival times, quantum vacuum fluctuations, phase noise, or spin states. These events are governed by the laws of physics rather than deterministic processes. This makes QRNGs fundamentally immune to the reverse engineering or prediction attacks.

In post-quantum cryptographic scenarios, the need for such true randomness becomes even more critical. For example, if the entropy source used to generate keys or nonces is compromised, even the most robust lattice, or code-base schemes could become susceptible to side-channel or cryptanalytic attacks. QRNGs not only eliminate this vulnerability but also enable high-assurance entropy provisioning [140]. This is essential for secure key generation, rekeying protocols, ephemeral key exchanges, and token signing in post-quantum protocols.

Moreover, in a distributed IoT network, ensuring that entropy sources are both independent and resistant to adversarial manipulation is critical. QRNGs provide a hardware-based assurance that each node generates cryptographic parameters without external influence or correlation. This hardware-based unpredictability enhances the network's resilience to coordinated or simultaneous entropy attacks.

#### 4.6.4.2 Applications in Nonce Generation, Session Keys, and Secure Boot

The application of QRNGs in real-world cryptographic workflows for the IoT spans a variety of domains. As quantum-safe protocols increasingly rely on ephemeral and high-entropy keys, the ability to embed robust randomness into various operational layers of an IoT device becomes essential. The three most prominent areas of QRNG application are nonce generation, session key derivation, and secure boot sequences.

*Nonce Generation*: Nonces are single-use random values that protect protocols from replay attacks, ensure message uniqueness, and contribute to the unpredictability of cryptographic exchanges. In many IoT devices, however, system entropy is weak or reused across boot cycles. These lead to nonce collisions that can compromise authentication or message integrity. By integrating QRNGs, devices can ensure every nonce used in TLS [50], DTLS [31], CoAP [30], or device pairing processes is drawn from a physically unpredictable quantum source.

*Session Key Derivation*: Post-quantum cryptographic protocols, such as those using lattice-based key encapsulation mechanisms (e.g., Kyber) [66, 71], rely heavily on secure ephemeral key generation. QRNGs can seed these key derivation functions with high-quality entropy. This improves the strength of key exchange and session confidentiality.

*Secure Boot and Firmware Verification*: Secure boot mechanisms depend on the integrity of the firmware image and the authenticity of its source. QRNGs can be used during the boot process to generate one-time tokens or challenges used in mutual authentication with the trusted firmware provider. They also serve as entropy sources for generating cryptographic checksums or signing keys that change with every device restart.

Beyond these core use cases, QRNGs can support dynamic key rotation, randomized memory address mapping to counter memory-based exploits, and generation of unique device fingerprints. These advanced functions are increasingly relevant as adversaries begin to combine quantum-assisted computation with conventional hardware attacks.

### 4.6.5 Quantum Key Distribution (QKD) in IoT Networks

QKD enables the secure exchange of cryptographic keys using the principles of quantum mechanics, such as the uncertainty principle and quantum entanglement [96]. Unlike traditional key exchange mechanisms that rely on computational complexity assumptions (e.g., RSA or ECC), QKD guarantees key secrecy based on the laws of physics. Any attempt at eavesdropping on a quantum channel introduces detectable disturbances, alerting legitimate users to a breach attempt. This property makes QKD particularly attractive in a post-quantum world.

The integration of QKD into IoT ecosystems is of growing interest due to the sensitivity and volume of data transmitted across interconnected smart devices [141]. IoT systems, ranging from wearable health monitors to autonomous vehicles, often require lightweight, scalable, and secure communication protocols. QKD offers a promising paradigm shift by ensuring unconditionally secure key exchanges that can form the foundation of encrypted IoT communication frameworks. However, applying QKD in IoT environments involves several architectural, logistical, and technological challenges. These are discussed in Section 4.6.6.1.

### 4.6.5.1 Architectural Challenges: Fiber vs Free-Space Implementation

There are two predominant physical-layer approaches to QKD: fiber-based and *free-space optical* (FSO) communication. Fiber-based QKD systems use optical fiber networks to transmit quantum signals, offering high stability and noise resistance [142]. These are suitable for metropolitan areas with dense infrastructure, such as smart city control centers and industrial IoT networks. However, the requirement for dedicated optical fibers restricts flexibility and mobility.

On the other hand, free-space QKD systems enable the transmission of qubits through atmospheric channels, such as air or space [143]. This makes FSO QKD more adaptable to dynamic IoT settings like drone fleets, autonomous vehicle corridors, or field-deployed sensor arrays. The downside is susceptibility to line-of-sight disruptions, atmospheric turbulence, weather conditions, and alignment drift, which compromise reliability and distance.

For low-power IoT devices, integrating full QKD transceivers remains impractical due to their size, power consumption, and cost. This has led to research into edge-based QKD models, where secure key exchange is managed by trusted intermediate nodes (e.g., routers), which in turn distribute session keys to IoT endpoints using lightweight encryption schemes [144-146].

### 4.6.5.2 Use Cases in IoT Environments

Several use-cases in IoT with QKD-enabled security can be imagined: (i) smart homes, (ii) medical wearables, (iii) vehicular networks, (iv) Industrial IoT (IIoT).

*Smart Homes*: QKD-enabled home gateways can perform secure key exchanges with cloud servers or ISPs and relay session keys to connected devices such as thermostats, voice assistants, and surveillance cameras [141]. This helps ensure encrypted communications with the devices.

*Medical Wearables*: Devices such as continuous glucose monitors and cardiac telemetry sensors transmit sensitive data. QKD can secure these transmissions, especially when aggregated through hospital-controlled IoT hubs that manage key generation and exchange.

*Vehicular Networks*: Vehicle-to-vehicle (V2V) and vehicle-to-infrastructure (V2I) communications are highly latency-sensitive and privacy-critical. Roadside units or mobile QKD terminals integrated into traffic lights can manage secure communications with vehicles [142].

*Industrial IoT (IIoT)*: Manufacturing and critical infrastructure deployments benefits from QKD-secured SCADA systems. Central QKD nodes within plants can distribute keys for encrypting control commands and sensor telemetry [147].

These use cases show that QKD can be selectively applied to secure high-value communication layers in IoT systems while using classical cryptography for less critical nodes.

### 4.6.5.3 Trusted Node vs. Device-Independent QKD and Scalability

The trusted node model is the current industry standard for QKD deployment. In this model, intermediary nodes generate or relay keys and are assumed to be secure. This works well in enterprise or city-wide networks but introduces a crucial point of vulnerability. If a trusted node is compromised, whether by cyberattack or insider threat, the confidentiality of the keys it relayed is also compromised.

*Device-Independent QKD (DI-QKD)* addresses this issue by removing the need to trust the hardware or internal workings of the QKD devices themselves. DI-QKD relies on observed violations of Bell inequalities, a concept from quantum non-locality, to confirm that key generation is secure even if the devices are faulty or malicious [148]. Although DI-QKD represents the gold standard of security, it is still largely experimental [149]. The key generation rates are low, the setup is complex, and transmission distances are short. Widespread adoption in IoT networks will require major advances in photonics and quantum information theory.

Scalability remains a core barrier to QKD adoption in the IoT. Deploying QKD for millions of endpoints is infeasible under current models. Hybrid architecture, where QKD is used for key generation at high-value aggregation points and classical post-quantum encryption is used at the leaf nodes, provide a more viable path forward. Standardization efforts, such as those led by ETSI and ITU-T, are working on interoperable protocols to bridge QKD and classical key distribution in a scalable manner [111, 150].

The emergence of satellite-based QKD, already demonstrated by China's Micius satellite, opens the possibility of global-scale QKD [151]. These systems could eventually provide secure keys to base stations that serve IoT clusters in rural or hard-to-wire environments, bypassing the limitations of terrestrial fiber infrastructure.

While widespread adoption of QKD across all IoT layers is constrained by current technical and economic limitations, its strategic deployment in critical infrastructure, smart healthcare, and transportation systems holds significant promises.

### 4.6.6 Privacy-Preserving Protocols Using Quantum Techniques

As quantum computing advances, its role in enhancing privacy-preserving technologies becomes increasingly critical, especially within IoT domains. Quantum-enhanced privacy-preserving protocols leverage the unique properties of quantum mechanics, such as superposition, entanglement, and no-cloning, to protect data integrity and enable secure multi-device coordination in complex IoT ecosystems. This section elaborates on three emerging techniques: Quantum Private Information Retrieval (QPIR) [152], Quantum Homomorphic Encryption in Federated Learning (QHE-FL) [153], and Secure Multi-Party Quantum Computation (SMPQC) [154].

#### 4.6.6.1 Quantum Private Information Retrieval (QPIR)

Private Information Retrieval (PIR) allows a client to retrieve records from databases without revealing their queries. Classical PIR protocols typically require either multiple non-colluding servers or introduce significant computational and communication overhead. In contrast, Quantum PIR (QPIR) exploits quantum properties like superposition to encode multiple query possibilities simultaneously. In a typical QPIR setup, the client prepares a quantum superposition over all database indices and sends the quantum query to a quantum-capable server [155]. The server evaluates the query on its quantum memory and returns a transformed quantum state, from which the client can extract the desired record without revealing which entry was accessed. This level of anonymity is particularly valuable in IoT environments like:

Recent studies have demonstrated that QPIR can achieve sub-linear communication complexity [152]. This outperforms classical PIR even under stringent constraints.

### 4.6.6.2 Quantum Homomorphic Encryption and Federated Learning

*Homomorphic Encryption* (HE) allows computations on encrypted data without requiring decryption, thereby preserving privacy during processing [156]. *Quantum Homomorphic Encryption* (QHE) extends classical homomorphic encryption by allowing quantum computations to be performed on encrypted quantum data without revealing the plaintext [157]. This is vital in decentralized and federated environments where IoT nodes participate in joint learning models without compromising local data. In decentralized IoT systems, where edge devices contribute to collective intelligence through federated learning, QHE serves as a foundational privacy layer. Each IoT device locally trains a machine learning model using encrypted sensor data and sends encrypted gradients to a central aggregator. QHE ensures that the aggregator cannot reverse-engineer raw data or individual model weights, preserving local privacy.

Despite challenges related to noise, circuit depth, and decoherence, variational QHE circuits are now being evaluated on *Noisy Intermediate-Scale Quantum* (NISQ) devices [158]. Hybrid quantum-classical federated learning frameworks are also under active development, combining the strengths of classical pre-processing with quantum secure aggregation.

### 4.6.6.3 Smart Data Aggregation and Multi-Party Quantum Computation

Secure Multi-Party Computation (SMPC) allows multiple parties to compute a joint function over their inputs without disclosing them to each other [159]. In its quantum counterpart, Secure Multi-Party Quantum Computation (SMPQC), entangled qubits and teleportation protocols are used to facilitate joint computations across distributed quantum nodes [160].

In IoT context, SMPQC facilitates collaborative analysis among multiple devices while preserving data locality and user privacy. For example, multiple environmental sensors in a smart building can use quantum resources to aggregate temperature or pollution data without revealing individual readings. Similarly, connected vehicles in a city can use SMPQC protocols to collaboratively detect traffic anomalies while protecting the location and route history of individual vehicles. Hospitals may compute disease patterns collaboratively without exposing patient-level data across institutions.

The advantage of SMPQC lies not only in privacy but also in resilience. Quantum channels inherently detect eavesdropping attempts through no-cloning theorem. Moreover, data aggregation over quantum networks can be performed at edge nodes that perform post-processing using hybrid algorithms.

Fig 4.14 illustrates key quantum techniques for privacy preservation in IoT ecosystems, including QPIR, QHE in federated learning, and SMQPC for secure distributed analytics.

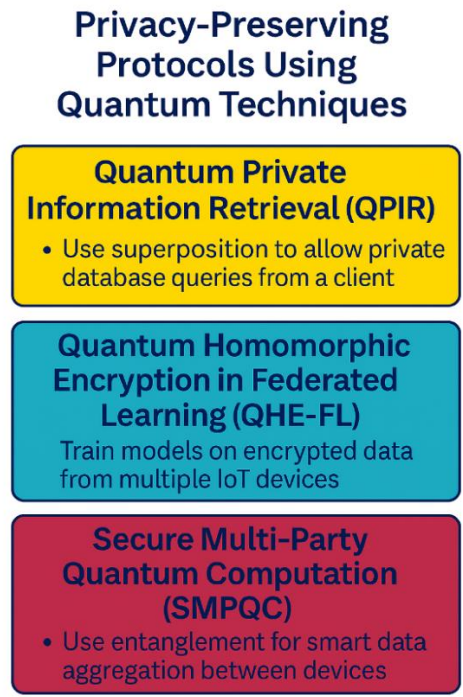

**Fig 4.14**: Quantum privacy-preserving protocols for IoT, including QPIR, QHE-FL, and SMPQC.

### 4.6.7 Privacy Governance and Regulatory Compliance

The adoption of quantum-resistant cryptographic techniques in IoT systems is not only a technical imperative but also a regulatory necessity. As privacy becomes increasingly critical, compliance with legal frameworks such as the General Data Protection Regulation (GDPR) [110], the Health Insurance Portability and Accountability Act (HIPAA) [130], and emerging global data protection laws [131] become increasingly challenging in the face of quantum threats. This section explores how privacy governance must evolve to account for the quantum landscape.

GDPR and HIPAA among the most stringent data protection frameworks globally, require that personal data be processed securely and remain confidential over its lifecycle. However, quantum computing introduces new risks under the *harvest now, decrypt later* paradigm [13]. Data encrypted using classical cryptographic schemes may be intercepted and stored today, only to be decrypted in the future using quantum techniques, potentially violating long-term confidentiality requirements set forth by GDPR Article 5 and HIPAA's Security Rule. This has led to the interpretation that forward secrecy and quantum-resistant encryption may soon become mandatory under regulatory best practices.

To address these emerging concerns, regulatory bodies are actively initiating efforts to support post-quantum transitions. NIST has finalized the selection of PQC algorithms such as CRYSTALS-Kyber [66, 71] and Dilithium [72] for public key encryption and digital signatures.

The European Union Agency for Cybersecurity (ENISA) has recommended a roadmap for critical infrastructure operators to migrate toward quantum-resilient architectures [161]. These initiatives signal a regulatory shift that places cryptographic agility and long-term privacy preservation at the center of digital governance.

In PQC-enabled systems, privacy governance also involves redefining trust relationships. Traditional Public Key Infrastructure (PKI) models rely on centralized certificate authorities and fixed public key lifetimes. Quantum threats challenge these models by rendering long-term keys insecure. Therefore, trust management must evolve to support ephemeral key exchanges, lattice-based digital signatures, and potentially, blockchain-based certificate transparency mechanisms. In smart cities and large-scale IoT ecosystems, decentralized identity frameworks integrated with PQC can provide robust authentication while minimizing the risk of key compromise.

Furthermore, national and regional governments are beginning to propose legislation that acknowledges the urgency of quantum readiness. For example, the U.S. Quantum Computing Cybersecurity Preparedness Act (QCCPA) mandates federal agencies to identify cryptographic systems vulnerable to quantum attacks and begin migration planning [162]. Similar initiatives are underway in countries like Japan, Canada, and Germany, indicating a convergence of regulatory priorities on global quantum cybersecurity resilience.

### 4.6.8 Standardization and Interoperability in Post-Quantum IoT Security

The transition to PQC systems in IoT infrastructures depends not only on technological innovation but also on coordinated standardization efforts and cross-platform interoperability. As quantum threats become closer to reality, ensuring that security mechanisms are harmonized across hardware, software, protocols, and jurisdictions become essential. This section explores the roles of major international bodies in PQC and QKD standardization, the implications for key communication protocols used in the IoT, and the challenges of achieving seamless integration across vendors and borders.

Several global organizations are at the forefront of PQC and QKD standardization. NIST in the United States is leading the most prominent initiative through its PQC standardization project, which began in 2016. As of 2024, NIST has selected CRYSTALS-Kyber [66, 71] for public-key encryption and CRYSTALS-Dilithium [72] for digital signatures as the core PQC standards. These selections are already influencing implementation choices in embedded IoT systems due to their performance and resistance to known quantum attacks.

In Europe, the European Telecommunications Standards Institute (ETSI) has established the Industry Specification Group for QKD (ISG-QKD) [111]. This group defines architecture models, key management protocols, and interface specifications suitable for quantum-enhanced security. ETSI's initiatives include testbeds and interoperability plugfests where multiple vendors evaluate their QKD and PQC solutions against reference frameworks.

The International Telecommunication Union - Telecommunication Standardization Sector (ITU-T) has contributed to defining quantum communication standards under Study Group 17 [163]. Their QKD specifications focus on trusted repeater architectures, inter-domain compatibility, and resilience to fault injection or side-channel exploitation. The Internet Engineering Task Force (IETF) has proposed hybrid key exchange mechanisms for TLS that combine traditional public-key cryptography with PQC primitives [164].

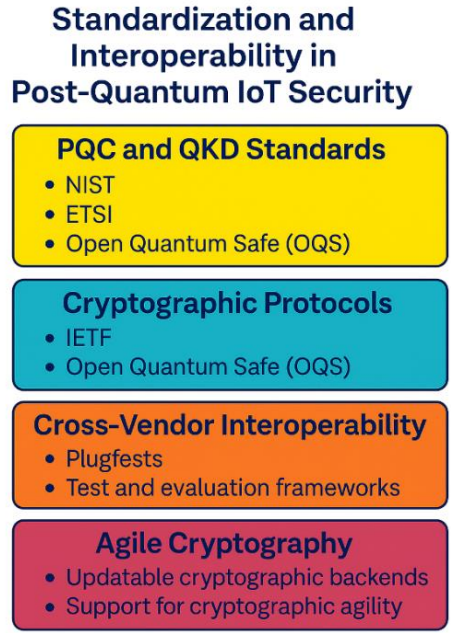

**Fig 4.15**: Standardization and interoperability framework for post-quantum IoT security.

The heterogeneity in IoT deployments demands standardized APIs and cryptographic libraries. To this end, the Open Quantum Safe (OQS) project provides open-source implementations of NIST candidates and APIs to ease integration in protocol stacks [165]. Similarly, IoT Security Foundation and GlobalPlatform are developing compliance frameworks that address the needs of quantum-safe device certification. Further, interoperability must be maintained across geopolitical boundaries. To address this, international bodies such as ISO/IEC JTC 1 are working toward harmonizing post-quantum standards across nations [166]. Their efforts support the portability of PQC libraries and certification schemes across jurisdictions. In addition to technical and legal concerns, operational interoperability requires support for cryptographic agility. Cryptographic agility refers to a system's ability to adopt, switch, or update cryptographic primitives with minimal disruption. For IoT devices, this means building firmware that can support pluggable cryptographic backends or deploying updatable secure elements that can accommodate future PQC upgrades. Open standards like CMS (Cryptographic Message Syntax) and CBOR Object Signing and Encryption (COSE) are being revised to support agile cryptography. Test and evaluation frameworks are also emerging as critical components for interoperability assurance. Initiatives such as NIST's National Cybersecurity Center of Excellence (NCCoE) provide PQC migration labs where vendors can test protocol performance and compatibility [167]. In Europe, quantum security interoperability pilots under the Horizon Europe program aim to evaluate QKD network integration in real-world 5G and IoT environments.

Fig 4.15 exhibits the collaborative standardization and interoperability initiatives shaping secure and privacy-preserving post-quantum IoT deployment across international and industry-led efforts. The combined efforts of international standardization bodies, open-source initiatives, and industry consortia are laying the groundwork for a secure, cross-compatible post-quantum IoT ecosystem.

### 4.6.9 Case Studies and Emerging Architectures

As PQC research transitions into real-world applications, pilot deployments and architectural prototypes are beginning to showcase how quantum-resistant technologies can be integrated into IoT infrastructures. This section presents case studies from major quantum research consortia and introduces architecture models that combine PQC, QRNGs, and cloud-edge paradigms.

#### 4.6.9.1 Pilot Deployments of PQC in the IoT

*OpenQKD*: Funded by the European Commission, OpenQKD is one of the most ambitious testbed projects exploring the integration of QKD across smart cities, critical infrastructure, and telecom networks [117]. The project has demonstrated how QKD can secure sensor-to-cloud links in IoT contexts such as healthcare and power grids. In Vienna, QKD-secured smart meters and surveillance feeds are being evaluated for resilience against eavesdropping and tampering.

*Quantum Internet Alliance (QIA)*: Coordinated by Delft University of Technology, the QIA aims to build a European quantum internet that interconnects quantum devices and classical infrastructure [168]. Pilot nodes within the alliance have begun exploring how PQC algorithms can support secure message routing and device authentication for IoT gateways and edge processors.

*China's Quantum Network Pilot Zones*: Several urban zones in China are now linked by metropolitan quantum networks using fiber-optic QKD [106]. These deployments include smart traffic control, encrypted government IoT communications, and health monitoring applications secured by PQC and QKD hybrids.

*Swiss Quantum Initiative*: In Switzerland, experimental quantum-safe communication channels are being evaluated across critical infrastructure including public transportation and emergency response networks [107]. These use combinations of PQC algorithms and centralized QRNG entropy servers to test zero-downtime, zero-trust communication layers.

### 4.7 Conclusion

The chapter has explored the intersection of quantum computing and Internet of Things (IoT) security, highlighting the profound challenges and opportunities that arise in this evolving landscape. Beginning with an overview of classical cryptographic techniques such as RSA, ECC, and AES, the discussion showed how these algorithms currently form the foundation of IoT security. However, advances in quantum algorithms like Shor's and Grover's pose direct threats to these systems, rendering them increasingly vulnerable in the era of quantum computing. To address these concerns, the chapter introduced post-quantum cryptography (PQC) families, such as lattice-based, code-based, hash-based, and multivariate schemes, and assessed their suitability for IoT environments that are resource-constrained. In addition, the potential of quantum-based techniques, particularly QKD and QRNGs, was examined in the context of securing communication in complex and large-scale IoT infrastructure such as smart cities. By presenting the underlying principles, operational flows, and standardization initiatives, the chapter has provided a comprehensive, tutorial-style overview that can serve as a guide for researchers, students, and practitioners seeking to understand the implications of quantum technologies for IoT security and privacy.

Looking ahead, future work in this domain is expected to focus on bridging the gap between theoretical advancements and practical deployments of quantum-resilient security. While PQC algorithms are progressing toward standardization, their adaptation to lightweight, energy-efficient implementations remains an open challenge for IoT ecosystems. Research into optimizing cryptographic primitives for constrained devices, along with cryptographic agility that allows seamless migration across algorithms, will be crucial. Similarly, quantum cryptographic solutions such as QKD must evolve to achieve scalability, cost-effectiveness, and interoperability with classical networks before they can be widely adopted in IoT infrastructures. Another promising avenue lies in hybrid architectures that combine the strengths of PQC and quantum cryptography, ensuring robust protection across diverse application domains. In parallel, interdisciplinary research involving security, systems engineering, and regulatory frameworks will be vital to address privacy concerns and governance in quantum-enhanced IoT. As the field matures, collaborative efforts between academia, industry, and standard bodies will play a central role in shaping secure and resilient IoT ecosystems capable of withstanding the disruptive impact of quantum computing.